\documentclass[a4paper, 11pt]{article}
\pdfoutput=1

\usepackage[utf8]{inputenc}

\usepackage{amsmath,amsfonts,amstext,amssymb,mathrsfs,amsthm}
\usepackage{jheppub2020}
\usepackage{graphicx}
\usepackage{bm}
\usepackage{ifthen}
\usepackage[all]{xy}
\usepackage{xcolor}
\usepackage{comment}
\usepackage[numbers,sort&compress]{natbib}
\usepackage{tikz}

\renewcommand{\i}{\mathcal{I}}

\newcommand{\be}{\begin{equation}}
\newcommand{\ee}{\end{equation}}
\renewcommand{\[}{\begin{equation}}
\renewcommand{\]}{\end{equation}}

\newcommand{\I}{\mathrm{i}}
\newcommand{\D}{\mathrm{d}}

\newcommand{\Z}{\mathbb{Z}}

\renewcommand{\O}{\mathcal{O}}

\newcommand{\<}{\langle}
\renewcommand{\>}{\rangle}

\newcommand{\nn}{\nonumber}
\newcommand{\lla}{\langle \! \langle}
\newcommand{\rra}{\rangle \! \rangle}

\newcommand{\bs}[1]{\boldsymbol{#1}}
\newcommand{\x}{{\bs{x}}}

\newcommand{\p}{\bs{p}}
\newcommand{\q}{{\bs{q}}}

\DeclareMathOperator{\K}{\mathcal{K}}

\title{Conformal correlators as simplex integrals \\ in momentum space}

\author[a]{Adam Bzowski,}
\affiliation[a]{Department of Physics and Astronomy, 
Uppsala University,
751 08 Uppsala, Sweden.}

\author[b]{Paul McFadden}
\affiliation[b]{School of Mathematics, 
Statistics \& Physics, Newcastle University, 
Newcastle NE1 7RU, U.K.} 

\author[c]{and Kostas Skenderis.}
\affiliation[c]{STAG Research Center \& Mathematical Sciences, 
University of Southampton,
Highfield, \\
Southampton 
SO17 1BJ, U.K.}

\emailAdd{adam.bzowski@physics.uu.se}
\emailAdd{paul.l.mcfadden@newcastle.ac.uk}
\emailAdd{k.skenderis@soton.ac.uk}

\begin{document}

\abstract{

We find the general solution of the conformal Ward identities for scalar $n$-point functions in momentum space and in general dimension. The solution is given in terms of integrals over $(n-1)$-simplices in momentum space. The $n$ operators are inserted at the $n$ vertices of the simplex, and the momenta running between any two vertices of the simplex are the integration variables.  The  integrand involves an arbitrary function of momentum-space cross ratios constructed from the integration variables, while the external momenta enter only via momentum conservation at each vertex. Correlators where the function of cross ratios is a monomial exhibit a remarkable recursive structure where $n$-point functions are built in terms of $(n-1)$-point functions. To illustrate our discussion, we derive the simplex representation of $n$-point contact Witten diagrams in a holographic conformal field theory.  This can be achieved through both a recursive method, as well as an approach based on the star-mesh transformation of electrical circuit theory.
The resulting expression for the function of cross ratios involves $(n-2)$ integrations, which is an improvement (when $n>4$) relative to the Mellin representation that involves $n(n-3)/2$ integrations.

}

\maketitle

\section{Introduction}

The general form of position-space $n$-point correlators in a conformal field theory (CFT) has been known for half a century \cite{Polyakov:1970xd}.  In \cite{Bzowski:2019kwd}, we presented a counterpart for this result in momentum space: 
a representation for the general momentum-space $n$-point function as a Feynman integral over an $(n-1)$-simplex, featuring an arbitrary function of the  cross ratios constructed from the momenta running between the vertices.  These cross ratios play an equivalent role to those in position space, and ensure the correlator has the same number of degrees of freedom in either basis.
In this paper, we prove the conformal invariance of this simplex representation, for arbitrary $n$ and in any spacetime dimension.
We then apply our results to a prototypical class of CFT correlators: $n$-point contact Witten diagrams. 
This illustrates two themes of a general nature: first, the applicability of results from electrical circuit theory relating resistor networks of different topologies; and second, the power of recursive methods when used in combination with the simplex representation.

Despite the late start, there are by now many areas of research where a knowledge of momentum-space CFT correlators is of practical value. 
Perhaps most noteworthy is the study of inflationary correlators, 
where the de Sitter isometries act on late-time slices as conformal transformations. (For a sample of works from a holographic perspective, see \cite{Witten:2001kn, Strominger:2001pn, Strominger:2001gp, Maldacena:2002vr,  Antoniadis:2011ib, Maldacena:2011nz,  Anninos:2011ui,Bzowski:2011ab, Kehagias:2012pd, Bzowski:2012ih,Mata:2012bx, McFadden:2013ria,  Ghosh:2014kba, Anninos:2014lwa,Kundu:2014gxa,Garriga:2014fda, Isono:2016yyj,Anninos:2019nib}, and \cite{ Arkani-Hamed:2015bza,Arkani-Hamed:2018kmz,Baumann:2019oyu,Baumann:2020dch,Sleight:2019mgd,Sleight:2019hfp, Sleight:2020obc} for the related cosmological collider and bootstrap perspective.\footnote{We emphasise too that 
slow roll can be understood through conformal perturbation theory \cite{Bzowski:2012ih,McFadden:2013ria}.})
Indeed, going beyond the many specific examples analysed in these and other works, our results here can be interpreted as providing the {\it most general} form of late-time momentum-space $n$-point correlators consistent with de Sitter symmetry. 

Other interesting applications include condensed matter physics, particularly quantum critical transport \cite{Chowdhury:2012km,Huh:2013vga, Jacobs:2015fiv,  Lucas:2016fju, Lucas:2017dqa} and anomalous hydrodynamics \cite{Chernodub:2019tsx}; particle phenomenology \cite{Armillis:2009pq,Coriano:2012wp,Coriano:2018bbe} and anomalies \cite{Coriano:2017mux,Coriano:2018zdo}, and revealing double-copy structure inherited from scattering amplitudes \cite{Farrow:2018yni, Lipstein:2019mpu}.  Momentum-space methods are moreover useful for the analysis of renormalisation \cite{Bzowski:2015pba, Bzowski:2017poo,Bzowski:2018fql}, where 
they offer a simple extraction of divergences plus an elegant decomposition of tensorial structure.  Looking ahead, a particularly exciting future application -- 
for which the present work is a necessary first step -- 
is the formulation of momentum-space approaches to the conformal bootstrap (see \cite{Simmons-Duffin:2016gjk, Poland:2018epd} for reviews and \cite{Isono:2018rrb, Isono:2019ihz, Gillioz:2018mto,Gillioz:2019iye,Sleight:2019ive} for related ideas).
With the form of general $n$-point correlators to hand,
understanding the partial wave decomposition and singularity structure of correlators is now  within reach.

The layout of this paper is as follows.  In section \ref{simpsec}, we present the simplex representation for general conformal correlators.   
In section \ref{meshsec}, we introduce 
a related class of `mesh' integrals with a remarkable recursive structure.  In section \ref{confinvsec}, we prove the conformal invariance of the simplex integral using three different methods.
In section \ref{holsec}, we derive the simplex representation for $n$-point contact Witten diagrams.  
Two independent methods are developed:
one using the star-mesh duality of electrical circuits, and the other based on a recursive application of the convolution theorem.  We conclude in section \ref{Discussionsec}.  
Appendix \ref{app:confproofs} presents technical aspects of    
our proofs of conformal invariance; appendix \ref{Symanzik_trick_appendix} reviews the Symanzik trick for  conformal integrals; and appendix \ref{app_Dfn} derives a new representation for the position-space holographic $D$-function in terms of a triple-$K$ integral.

\section{Conformal correlators as simplex integrals}
\label{simpsec}

\subsection{Position space}

In position space, any conformally invariant $n$-point function of scalar operators $\O_1, \ldots, \O_n$, of scaling dimensions $\Delta_1, \ldots, \Delta_n$, takes the well-known form \cite{Polyakov:1970xd, DiFrancesco:1997nk}
\begin{equation} \label{xcorr}
\< \O_1 (\bs{x}_1) \ldots \O_n(\bs{x}_n) \> = \prod_{1 \leq i < j \leq n} x_{ij}^{2 \alpha_{ij}} f(\bs{u}),
\end{equation}
where the parameters $\alpha_{ij}$ are related to the scaling dimensions by the relations
\begin{equation} \label{deltaijcond_a}
\Delta_m = - \sum_{j=1}^n \alpha_{mj}, \qquad m=1,2,\ldots,n.
\end{equation}
Without loss of generality we assume  $\alpha_{ji} = \alpha_{ij}$ and $\alpha_{ii} = 0$. We denote  the full set of parameters collectively as $\bs{\alpha} = \{ \alpha_{ij} \}_{1 \leq i < j \leq n}$. We assume Euclidean signature throughout. The function $f$ is an arbitrary function of the   conformal cross ratios,
\begin{equation}
u_{[pqrs]} = \frac{x_{pr}^2 x_{qs}^2}{x_{pq}^2 x_{rs}^2},
\end{equation}
where $p,q,r,s = 1,2,\ldots,n$ are distinct numbers.\footnote{Cross ratios are parametrised by four numbers; to distinguish these from other indices, we place them inside square brackets. No antisymmetrisation is implied.} However, this considerably over-counts their number: 
at most only $n(n-3)/2$  are independent.\footnote{In sufficiently high spacetime dimensions all $n(n-3)/2$  are independent, but in low spacetime dimensions there are degeneracies.}
There are various ways of choosing an independent set, but here we will use
\begin{align}
& u_{2a} = u_{[123a]} = \frac{x_{2a}^2 x_{13}^2}{x_{1a}^2 x_{23}^2}, && u_{3a} = u_{[132a]} = \frac{x_{3a}^2 x_{12}^2}{x_{1a}^2 x_{23}^2}, && u_{ab} = u_{[2a3b]} = \frac{x_{ab}^2 x_{23}^2}{x_{2a}^2 x_{3b}^2},
\end{align}
where $a, b = 4,\,5,\ldots,n$ and $a < b$, so we have $2(n-3)+(n-3)(n-4)/2=n(n-3)/2$ ratios in total. We  denote these independent cross ratios collectively as a vector $\bs{u}$. 

\subsection{Momentum space}

In \cite{Bzowski:2019kwd}, we showed the general scalar $n$-point function in momentum space can be expressed as a \emph{simplex integral}. This is defined as a Feynman integral over an oriented $\vphantom{\big[}(n-1)$-simplex,
\begin{align} \label{simplex}
&\< \O_1(\bs{p}_1) \ldots \O_n(\bs{p}_n) \> =
 \prod_{1 \leq i < j \leq n} \int \frac{\D^d \bs{q}_{ij}}{(2 \pi)^d} \frac{\hat{f}(\hat{\bs{u}})}{q_{ij}^{2 \alpha_{ij} + d}}  \prod_{k=1}^n (2\pi)^d \delta \Big( \bs{p}_k + \sum_{l=1}^n \bs{q}_{lk} \Big),
\end{align}
where $d$ is the spacetime dimension and the orientation is assigned by numbering the vertices.  We associate a momentum $\bs{q}_{ij}$ to the edge running from vertex $i$ to $j$  so that $\bs{q}_{ij} = -\bs{q}_{ji}$  and $\bs{q}_{jj} = 0$.  We  thus have $n(n-1)/2$ integration variables which we choose to be the $\bs{q}_{ij}$ with $i < j$.
The parameters $\alpha_{ij}$ satisfy \eqref{deltaijcond_a}. 
The function $\hat{f}$ is an arbitrary function of the  independent \emph{momentum-space cross ratios}:
\[ \label{conf_ratio_q}
\hat{u}_{[pqrs]} = \frac{q_{pq}^2 q_{rs}^2}{q_{pr}^2 q_{qs}^2}.
\]
As in position space (for sufficiently high spacetime dimensions), only $n(n-3)/2$ of these cross ratios are independent. We will choose the set
\begin{align}
& \hat{u}_{2a} = \hat{u}_{[123a]} = \frac{q_{12}^2 q_{3a}^2}{q_{2a}^2 q_{13}^2}, && \hat{u}_{3a} = \hat{u}_{[132a]} = \frac{q_{13}^2 q_{2a}^2}{q_{3a}^2 q_{12}^2}, && \hat{u}_{ab} = \hat{u}_{[2a3b]} = \frac{q_{2a}^2 q_{3b}^2}{q_{ab}^2 q_{23}^2},
\end{align}
where $a, b = 4,\,5,\ldots,n$ and $a < b$.  We will denote the set of indices enumerating the independent cross ratios as $\mathcal{U}$, while the ratios themselves will be written collectively as $\hat{\bs{u}}$, thus $\hat{\bs{u}} = \{ \hat{\bs{u}}_I \}_{I \in \mathcal{U}}$.

In momentum space, the cross ratios are subject to integration inside the simplex integral \eqref{simplex}. Overall, we  have $n(n-1)/2$ integrals and $n-1$ delta functions. Setting one delta function aside for overall momentum conservation, we will employ the double-bracket notation
\begin{align}
\< \O_1(\bs{p}_1) \ldots \O_n(\bs{p}_n) \> = (2 \pi)^d \delta \Big( \sum_{i=1}^n \bs{p}_i \Big) \lla \O_1(\bs{p}_1) \ldots \O_n(\bs{p}_n) \rra.
\end{align}
We can now perform the integrals over the variables $\bs{q}_{in}$ for $i=1,2,\ldots, n-1$ in \eqref{simplex} to remove the remaining delta functions. This leaves us with $(n-1)(n-2)/2$ integrals still to perform, corresponding to the simplex integral 
\begin{align} \label{simplex_red}
\lla \O_1(\bs{p}_1) \ldots \O_n(\bs{p}_n) \rra = \prod_{1 \leq i < j \leq n-1} \int \frac{\D^d \bs{q}_{ij}}{(2 \pi)^d} \frac{\hat{f}(\hat{\bs{u}})}{\text{Den}_{n}(\bs{\alpha})}
\end{align}
where the denominator reads
\begin{align}
\text{Den}_{n}(\bs{\alpha}) & = \prod_{1 \leq i < j \leq n-1} q_{ij}^{2 \alpha_{ij} + d} \times \prod_{m=1}^{n-1} |\bs{l}_m - \bs{p}_m|^{2 \alpha_{mn} + d} \label{Den}
\end{align}
and $\bs{l}_m$ depends only on the remaining internal momenta,
\begin{align}
\bs{l}_m =- \bs{q}_{mn} + \bs{p}_m = \sum_{j=1}^{n-1} \bs{q}_{mj} = - \sum_{j=1}^{m-1} \bs{q}_{jm} + \sum_{j=m+1}^{n-1} \bs{q}_{mj}.
\end{align}
Notice that we have eliminated the momentum $\bs{p}_n$ and hence all the remaining $\bs{p}_1, \ldots, \bs{p}_{n-1}$ are independent (assuming we are in sufficiently high spacetime dimensions). All sums and products now extend only up to $n-1$. We will refer to the expression \eqref{simplex_red} as the \emph{reduced simplex integral}.

\subsection{1-, 2- and 3-point functions}\label{sec_123ptfns}

For illustration, 
let us consider the 1-, 2- and 3-point functions of operators $\O_1, \O_2, \O_3$ with dimensions $\Delta_1, \Delta_2, \Delta_3$. Since for $n < 4$ there are no cross ratios, the function $\hat{f}_n = c_n$ is constant and disappears from the simplex integrals \eqref{simplex} and \eqref{simplex_red}.

For $n = 1$, the condition \eqref{deltaijcond_a} implies $\Delta_1 = 0$, and thus $\O_1$ is proportional to the identity operator and the reduced simplex integral is $\lla \O_1(\bs{p}_1) \rra = c_1$. 

For $n = 2$, the simplex integral reduces to 
\begin{align}
\lla \O_1(\bs{p}_1) \O_2(\bs{p}_2) \rra = c_2 p_1^{-2 \alpha_{12} - d}
\end{align}
and the conditions \eqref{deltaijcond_a} imply
\[
\Delta_1 = \Delta_2 = -\alpha_{12}.
\]

The 3-point function is the lowest correlator for which a loop integral survives,
\begin{align}\label{triint}
& \lla \O_1(\bs{p}_1) \O_2(\bs{p}_2) \O_3(\bs{p}_3) \rra = 
c_3 \int\frac{ \D^d \bs{q}}{(2\pi)^d}\frac{1}{q^{2\alpha_{12}+d}|\bs{q}-\bs{p}_1|^{2\alpha_{13}+d}|\bs{q}+\bs{p}_2|^{2\alpha_{23}+d}}
\end{align}
while the conditions \eqref{deltaijcond_a} imply
\[
2\alpha_{12}=2\Delta_3-\Delta_t = -\Delta_1-\Delta_2+\Delta_3
\]
along with cyclic permutations. This 1-loop triangle integral is equivalent to an integral of three modified Bessel $K$ functions (a `triple-$K$ integral'), which evaluates to a specific linear combination of the double hypergeometric function Appell $F_4$ as shown in \cite{Bzowski:2013sza}.   We will revisit this calculation later in section \ref{startrisection}.

We have thus recovered the known momentum-space expressions of 1-, 2- and 3-point functions \cite{Bzowski:2013sza, Coriano:2013jba} starting from the reduced simplex \eqref{simplex_red}.

\subsection{4-point function}

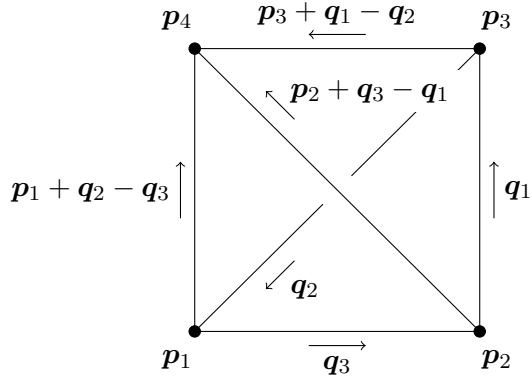
\begin{figure}[t]
\hspace{-1.6cm}	
\begin{tikzpicture}[scale=3.75]
		\draw (0,0) -- (0,1) -- (1,1) -- (1,0) -- cycle;
		\draw (0,0) -- (0.45,0.45);
		\draw (1,1) -- (0.55,0.55);
		\draw (1,0) -- (0,1);
		\node[right] at (-0.15,-0.1) {$\bs{p}_1$};
		\node[right] at (-0.15,1.1) {$\bs{p}_4$};
		\node[left] at (1.15,-0.1) {$\bs{p}_2$};
		\node[left] at (1.15,1.1) {$\bs{p}_3$};
		\draw[->] (0.4,-0.05) -- (0.6,-0.05);
		\node[below] at (0.5,-0.05) {$\bs{q}_3$};
		\draw[->] (0.6, 1.05) -- (0.4, 1.05);
		\node[above] at (0.5, 1.05) {$\bs{p}_3 + \bs{q}_1 - \bs{q}_2$};
		\draw[->] (-0.05, 0.4) -- (-0.05, 0.6);
		\node[left] at (-0.05, 0.5) {$\bs{p}_1 + \bs{q}_2 - \bs{q}_3$};
		\draw[->] (1.05, 0.4) -- (1.05, 0.6);
		\node[right] at (1.05, 0.5) {$\bs{q}_1$};
		\draw[<-] (0.25, 0.15) -- (0.35, 0.25);
		\node[right] at (0.3, 0.15) {$\bs{q}_2$};
		\node[right,fill=white] at (0.3, 0.85) {$\bs{p}_2 + \bs{q}_3 - \bs{q}_1$};
		\draw[->] (0.35, 0.75) -- (0.25, 0.85);
		\draw[black,fill=black] (0,0) circle [radius=0.02];
		\draw[black,fill=black] (1,0) circle [radius=0.02];
		\draw[black,fill=black] (0,1) circle [radius=0.02];
		\draw[black,fill=black] (1,1) circle [radius=0.02];
	\end{tikzpicture}
\centering
\caption{Representation of the 4-point function as the 3-simplex or tetrahedral integral \eqref{kint4}.  Each of the internal lines represents a generalised propagator in \eqref{Den3}, and the dots denote operator insertions carrying ingoing momenta $\p_i$. 
	\label{fig:1}}
\end{figure}

For the 4-point function, the simplex representation has three independent integrals. These correspond to the three loops of the 3-simplex (or tetrahedron) as shown in Fig.~\ref{fig:1}.  We find 
\begin{align}
& \lla \O_{1}(\bs{p}_1) \O_{2}(\bs{p}_2) \O_{3}(\bs{p}_3) \O_{4}(\bs{p}_4) \rra  = \int \frac{\D^d \bs{q}_{1}}{(2 \pi)^d} \frac{\D^d \bs{q}_{2}}{(2 \pi)^d} \frac{\D^d \bs{q}_{3}}{(2 \pi)^d} \frac{\hat{f}(\hat{u}, \hat{v})}{\text{Den}_4(\bs{q}_{j}, \bs{p}_k)}, \label{kint4}
\end{align}
where the denominator 
\begin{align}
 \text{Den}_4(\bs{q}_{j}, \bs{p}_k) &=q_{3}^{2 \alpha_{12} + d} q_{2}^{2 \alpha_{13} + d}  q_{1}^{2 \alpha_{23} + d} 
 \nn\\& \quad \times 
|\bs{p}_1 + \bs{q}_{2} - \bs{q}_{3}|^{2 \alpha_{14} + d}
 |\bs{p}_2 + \bs{q}_{3} - \bs{q}_{1}|^{2 \alpha_{24} + d} 
|\bs{p}_3 + \bs{q}_{1} - \bs{q}_{2}|^{2 \alpha_{34} + d}. \label{Den3}
\end{align}
This parametrisation corresponds to setting 
\begin{equation}
\bs{q}_i = \frac{1}{2} \epsilon_{ijk} \bs{q}_{jk}, \qquad (i, j, k=1, 2, 3)
\end{equation}
in \eqref{simplex_red}. The arbitrary 
function $\hat{f}(\hat{u},\hat{v})$ is now a function of just two variables,
\begin{align}\label{uvhatdef}
\hat{u} & = \frac{q_{1}^2 |\bs{p}_1 + \bs{q}_{2} - \bs{q}_{3}|^2}{q_{2}^2 |\bs{p}_2 + \bs{q}_{3} - \bs{q}_{1}|^2},\qquad  
\hat{v} = \frac{q_{2}^2 |\bs{p}_2 + \bs{q}_{3} - \bs{q}_{1}|^2}{q_{3}^2 |\bs{p}_3 + \bs{q}_{1} - \bs{q}_{2}|^2},
\end{align}
whose role is analogous to that of the position-space cross ratios $u$ and $v$. 
Notice though that  they 
depend on the momenta $\bs{q}_j$ which are 
subject to integration in \eqref{kint4}.

\section{Mesh integrals}
\label{meshsec}

We define a {\it mesh integral} as the generalised Feynman integral with $n$ points and $n(n-1)/2$ generalised propagators so that every pair of points is connected. A generalised propagator takes the form $q^{-2 \alpha}$ for some power $\alpha$ which is not necessarily equal to one. The mesh integral $M_n(\bs{\alpha}; \bs{p}_1,\ldots,\bs{p}_n)$ is thus
\begin{align} \label{mesh}
M_n(\bs{\alpha}; \bs{p}_1,\ldots,\bs{p}_n) = \prod_{1 \leq i < j \leq n} C_{ij} \int \frac{\D^d \bs{q}_{ij}}{(2 \pi)^d} \frac{1}{q_{ij}^{2 \alpha_{ij} + d}}  \prod_{k=1}^n (2\pi)^d \delta \Big( \bs{p}_k + \sum_{l=1}^n \bs{q}_{lk} \Big),
\end{align}
where the coefficient 
\begin{equation}\label{Cij_def}
C_{ij} = \frac{\pi^{d/2} 2^{d + 2 \alpha_{ij}}}{\Gamma(-\alpha_{ij})} \Gamma \Big( \frac{d}{2} + \alpha_{ij} \Big).
\end{equation}
is included for convenience.   For $n=1$, we define the mesh integral as $M_1(\bs{p}_1)=(2\pi)^d\delta (\bs{p}_1)$. As above, there are $n(n-1)/2$ integration variables $\bs{q}_{ij}$ with $i < j$ (and we extend $\bs{q}_{ij}$  to any $i, j$ by $\bs{q}_{ij} = -\bs{q}_{ji}$). A mesh integral is therefore a simplex integral with $\hat{f} = 1$. However, we are not placing any restriction on the values of the parameters $\alpha_{ij}$ here.  

Just as for simplex integrals, we define the \emph{reduced mesh integrals} $\tilde{M}_n(\bs{\alpha}; \bs{p}_1, \ldots, \bs{p}_n)$ by pulling out the momentum-conserving delta function,
\begin{align} 
M_n(\bs{\alpha}; \bs{p}_1, \ldots, \bs{p}_n) = (2 \pi)^d \delta \Big( \sum_{i=1}^n \bs{p}_i \Big) \tilde{M}_n(\bs{\alpha}; \bs{p}_1, \ldots, \bs{p}_n).
\end{align}
Up to the factors of $C_{ij}$, the reduced mesh integrals are given by \eqref{simplex_red} with $\hat{f} = 1$, namely 
\begin{align} \label{mesh_red}
\tilde{M}_n(\bs{\alpha}; \bs{p}_1, \ldots, \bs{p}_n) = \prod_{1 \leq i < j \leq n-1} C_{ij} \int \frac{\D^d \bs{q}_{ij}}{(2 \pi)^d} \frac{1}{\text{Den}_{n}(\bs{\alpha})},
\end{align}
where the denominator is given by \eqref{Den}.

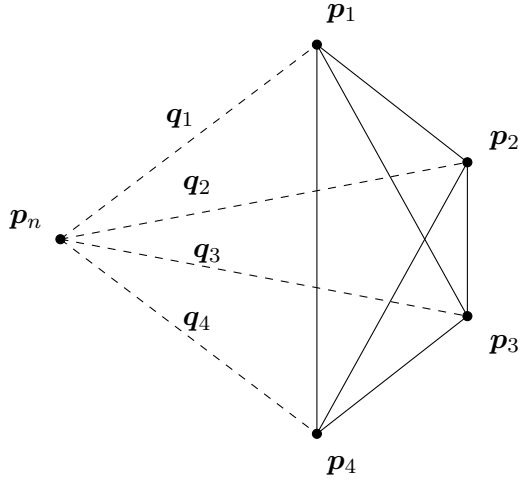
\begin{figure}[t]
\centering
\begin{tikzpicture}[scale=3,xscale=1.5]
\draw (0.5, 0.86) -- (0.94,0.34) -- (0.94,-0.34) -- (0.5, -0.86) -- cycle;
\draw (0.5, 0.86) -- (0.94,-0.34);
\draw (0.5, -0.86) -- (0.94,0.34);
\draw[dashed] (-0.25, 0) -- (0.5, 0.86);
\draw[dashed] (-0.25, 0) -- (0.94,0.34);
\draw[dashed] (-0.25, 0) -- (0.94,-0.34);
\draw[dashed] (-0.25, 0) -- (0.5, -0.86);
\node[right] at (0.5, 1.0) {$\bs{p}_1$};
\node[above] at (1.05, 0.35) {$\bs{p}_2$};
\node[above] at (1.05, -0.55) {$\bs{p}_3$};
\node[right] at (0.5, -1.0) {$\bs{p}_4$};
\node[above] at (-0.35, 0) {$\bs{p}_n$};
\node[above] at (0.1, 0.45) {$\bs{q}_1$};
\node[above] at (0.15, 0.15) {$\bs{q}_2$};
\node[above] at (0.18, -0.15) {$\bs{q}_3$};
\node[above] at (0.15, -0.45) {$\bs{q}_4$};
\draw[black,fill=black] (0.5, 0.86) circle [radius=0.02,xscale=0.66];
\draw[black,fill=black] (0.94,0.34) circle [radius=0.02,xscale=0.66];
\draw[black,fill=black] (0.94,-0.34) circle [radius=0.02,xscale=0.66];
\draw[black,fill=black] (0.5, -0.86) circle [radius=0.02,xscale=0.66];
\draw[black,fill=black] (-0.25, 0) circle [radius=0.02,xscale=0.66];
\end{tikzpicture}
\caption{The decomposition of the 5-point mesh $M_5$. The solid internal lines on the right-hand side of the figure represent the 4-point mesh $M_4$ evaluated with ingoing momenta $\bs{p}_j - \bs{q}_j$.\label{fig:2}}
\end{figure}

Mesh integrals exhibit a recursive structure. By pulling out factors containing $\q_{in}$ and $\bs{p}_n$ and renaming $\q_{in} \mapsto \q_{i}$, one can rewrite the mesh integral recursively as
\begin{align}\label{meshrec0}
&M_n(\bs{\alpha}; \bs{p}_1, \ldots, \bs{p}_n) \nn\\&\quad = \prod_{i = 1}^{n-1} C_{in} \int \frac{\D^d \bs{q}_{i}}{(2 \pi)^d} \frac{M_{n-1}(\bs{\alpha}; \bs{p}_1 - \bs{q}_1, \ldots, \bs{p}_{n-1} - \bs{q}_{n-1})}{q_1^{2 \alpha_{1n} + d} q_2^{2 \alpha_{2n} + d} \ldots q_{n-1}^{2 \alpha_{n-1,n} + d}} (2 \pi)^d \delta \Big( \bs{p}_n + \sum_{j=1}^{n-1} \bs{q}_j \Big).
\end{align}
This formula is depicted visually in figure \ref{fig:2} and will be used repeatedly in the analysis of the following section.

\subsection{Fourier transform}\label{ftmeshsec}

We now show that the mesh integral \eqref{mesh} is the Fourier transform of the position-space conformal $n$-point function \eqref{xcorr} when  $f$ is a {\it monomial} in the cross ratios, {\it i.e.,} a product of cross ratios raised to arbitrary powers.
For such an $f$, the correlator \eqref{xcorr}  takes the form
\begin{equation} \label{posF}
F_n(\bs{\alpha}; \bs{x}_1,\ldots,\bs{x}_n) = \prod_{1 \leq i < j \leq n} x_{ij}^{2 \alpha_{ij}},
\end{equation}
where the $\alpha_{ij}$ are still a solution of \eqref{deltaijcond_a}.
As we will see,  the Fourier transform of \eqref{posF}
is equal to \eqref{mesh}, namely 
\begin{align} \label{ft}
\mathcal{F}[F_n](\bs{\alpha}; \bs{p}_1, \ldots, \bs{p}_n) & = \int \D^d \bs{x}_1 \ldots \D^d \bs{x}_n \,e^{-\I \sum_{j=1}^n \bs{x}_j \cdot \bs{p}_j} F_n(\bs{\alpha}; \bs{x}_1, \ldots, \bs{x}_n) \nn\\
& = M_n(\bs{\alpha};\bs{p}_1,\ldots,\bs{p}_n).
\end{align}

Let us first check this for the 2-point function. The Fourier transform of $F_2$ is
\begin{equation} \label{ft2}
\mathcal{F}[x_{12}^{2 \alpha_{12}}] = (2 \pi)^d \delta(\bs{p}_1 + \bs{p}_2) \frac{C_{12}}{p_1^{2 \alpha_{12} + d}},
\end{equation}
where $C_{ij}$ is given in \eqref{Cij_def}. This matches the corresponding mesh, which from \eqref{mesh}, is
\begin{align}
M_2(\alpha_{12}; \bs{p}_1, \bs{p}_2) & = C_{12} \int \frac{\D^d \bs{q}_{12}}{(2 \pi)^d} \frac{1}{q_{12}^{2 \alpha_{12} + d}} (2 \pi)^d \delta(\bs{p}_1 - \bs{q}_{12}) (2 \pi)^d \delta(\bs{p}_2 + \bs{q}_{12}) \nn\\
& = (2 \pi)^d \delta(\bs{p}_1 + \bs{p}_2) \frac{C_{12}}{p_1^{2 \alpha_{12} + d}}.
\end{align}

We now proceed by induction and use the recursive structure expressed in \eqref{meshrec0}.  Let us assume the statement \eqref{ft} holds true up to the level of the $(n-1)$-point function. We can write $F_n$ as
\begin{equation}
F_n(\bs{\alpha}; \bs{x}_1,\ldots,\bs{x}_n) = x_{1n}^{2\alpha_{1n}} x_{2n}^{2\alpha_{2n}} \ldots x_{n-1,n}^{2\alpha_{n-1,n}} \times F_{n-1}(\bs{\alpha}; \bs{x}_1,\ldots,\bs{x}_{n-1}).
\end{equation}
Using the Fourier transform of a power function as given in \eqref{ft2}, and denoting convolution with a $\ast$, the Fourier transform is then 
\begin{align}\label{FTmesh}
\mathcal{F}[F_{n}] & = \mathcal{F} \left[ x_{1n}^{2\alpha_{1n}} x_{2n}^{2\alpha_{2n}} \ldots x_{n-1,n}^{2\alpha_{n-1,n}} \right] \ast \mathcal{F}[F_{n-1}]  \nn\\
& = \left[ \frac{(2 \pi)^d \delta \left( \sum_{j=1}^n \bs{p}_j \right)\prod_{i=1}^{n-1}C_{in}}{p_1^{2 \alpha_{1n} +d} p_2^{2 \alpha_{2n} +d} \ldots p_{n-1}^{2 \alpha_{n-1,n} +d}} \right] \ast \left[ M_{n-1}(\bs{\alpha}; \bs{p}_1, \ldots, \bs{p}_{n-1}) \, (2 \pi)^d \delta(\bs{p}_n) \right] \nn\\&=
\prod_{i = 1}^{n-1} C_{in} \int \frac{\D^d \bs{q}_{i}}{(2 \pi)^d} \frac{M_{n-1}(\bs{\alpha}; \bs{p}_1 - \bs{q}_1, \ldots, \bs{p}_{n-1} - \bs{q}_{n-1})}{q_1^{2 \alpha_{1n} + d} q_2^{2 \alpha_{2n} + d} \ldots q_{n-1}^{2 \alpha_{n-1,n} + d}} (2 \pi)^d \delta \Big( \bs{p}_n + \sum_{j=1}^{n-1} \bs{q}_j \Big)\nn\\
&= M_n(\bs{\alpha};\bs{p}_1,\ldots,\bs{p}_n)
\end{align}
where we used the mesh recursion relation  \eqref{meshrec0}.

\section{Conformal invariance}
\label{confinvsec}

In this section, we now show that the simplex integrals \eqref{simplex} are conformally invariant. 
The conformal invariance of the mesh integrals \eqref{mesh} is already apparent, since as we showed above they correspond to the Fourier transform of the   position-space CFT correlator \eqref{xcorr}  when $f$ is a monomial in the cross ratios.
As we discuss below, the conformal invariance of mesh integrals can also be proven directly using the momentum-space conformal Ward identities (CWIs), which provides a useful warm-up for the simplex.  In fact, as we will see, the conformal invariance of the simplex follows directly from the invariance of the mesh.

The layout of this section is as follows.  First, in section \ref{CWIs_sec} we  review the form of the momentum-space CWIs.  In section \ref{mesh_sec}, we then demonstrate that mesh integrals are a solution.  For the special conformal Ward identity (SCWI), this can be shown either recursively, or else by showing that the action of the corresponding differential operator on the mesh yields a total derivative.  To keep the discussion brief, we present only the key ideas and relegate all technical details to appendices \ref{app:conf_mesh_P} and \ref{app:conf_mesh_A}.
In section \ref{simpsubsec}, we then prove the conformal invariance of the simplex.  This can be done by three routes.  The first is to relate the simplex to the mesh via a Mellin-Barnes transform as discussed in section \ref{sec:conf_simplex_P}.  The other two routes involve showing that   the action of the SCWI operator on the simplex yields a total derivative. This can be accomplished either indirectly, as shown in section \ref{sec:conf_simplex_K}, or else by direct computation as shown in section \ref{sec:conf_simplex_A}.  Here we again present only the key ideas, leaving the details to appendix \ref{app:conf_simp_A}.

\subsection{Conformal Ward identities} \label{CWIs_sec}

In Euclidean signature, the conformal group
consists of  translations, rotations, dilatations and special conformal transformations \cite{DiFrancesco:1997nk}. Invariance under rotations requires  the simplex \eqref{simplex} and mesh \eqref{mesh} are built from invariant scalar products, which is indeed the case. Invariance under translations generates the overall momentum-conserving delta functions.
Invariance under dilatations and special conformal transformations can be expressed via the conformal Ward identities below. We will use the Ward identities for both the full correlators  as well as the reduced correlators  where the delta function of momentum conservation has been removed, which we write as $\<\ldots\>$ and $\lla \ldots\rra$ respectively. 

Setting $\Delta_t = \sum_{i=1}^n \Delta_i$, the dilatation Ward identity is
\begin{align} \label{dil}
\left[ \Delta_t - n d - \sum_{m=1}^{n-1} p_m^{\mu} \frac{\partial}{\partial p_m^\mu} \right] \< \O_1(\bs{p}_1) \ldots \O_n(\bs{p}_n) \> = 0,
\end{align}
or for the reduced correlator,
\begin{align}
\left[ \Delta_t - (n-1)d - \sum_{m=1}^{n-1} p_m^{\mu} \frac{\partial}{\partial p_m^\mu} \right] \lla \O_1(\bs{p}_1) \ldots \O_n(\bs{p}_n) \rra = 0. \label{dil_red}
\end{align}
The first of these equations fixes total dimension of the full correlator to be $\Delta_t - n d$.  After removing the delta function of momentum conservation, the reduced correlator then has  dimension $\Delta_t - (n-1) d$ as imposed by the second equation.

The special conformal Ward identities (SCWIs) are
\begin{align} \label{special}
& \mathcal{K}^{\kappa}(\bs{\Delta}) \< \O_1(\bs{p}_1) \ldots \O_n(\bs{p}_n) \> = 0, \\
& \tilde{\mathcal{K}}^{\kappa}(\bs{\Delta}) \lla \O_1(\bs{p}_1) \ldots \O_n(\bs{p}_n) \rra = 0, \label{special_red}
\end{align}
where the corresponding SCWI operators are
\begin{align}
\mathcal{K}^{\kappa}(\bs{\Delta}) = \sum_{m=1}^{n} \mathcal{K}^{\kappa}(\Delta_m; \bs{p}_m), \\
\tilde{\mathcal{K}}^{\kappa}(\bs{\Delta}) = \sum_{m=1}^{n-1} \mathcal{K}^{\kappa}(\Delta_m; \bs{p}_m),
\end{align}
with
\begin{align} \label{Kopdef}
\mathcal{K}^\kappa(\Delta;\bs{p}) = p^\kappa \frac{\partial}{\partial p^\mu} \frac{\partial}{\partial p_\mu} - 2 p^\mu \frac{\partial}{\partial p^\mu} \frac{\partial}{\partial p_\kappa}+2(\Delta-d)\frac{\partial}{\partial p_\kappa}. 
\end{align}
Here, $\bs{\Delta}$ denotes collectively the scaling dimensions of the operators involved.  Our task is now to show that both the mesh and the simplex integrals satisfy these differential equations given appropriate values of the scaling dimensions.

\subsection{Mesh integrals}
\label{mesh_sec}

First, we show that the 
 mesh integral \eqref{mesh} satisfies 
 \eqref{dil} and \eqref{special} with scaling dimensions subject to  \eqref{deltaijcond_a}. Notice this condition can be interpreted in two ways. Given \emph{any} set of parameters $\bs{\alpha}$, the mesh integral satisfies the  Ward identities with dimensions given by \eqref{deltaijcond_a}. Alternatively, if  the dimensions are fixed, then we can solve \eqref{deltaijcond_a} for the $\bs{\alpha}$.  For $n\ge 4$ more than one solution is possible:  in general, there is an  $n(n-1)/2 - n = n(n-3)/2$ parameter family of allowed values for the $\bs{\alpha}$.   
From the simplex point of view, this corresponds to choosing a monomial $\hat{f}(\hat{\bs{u}})$ where the $\bs{\alpha}$ parametrise the powers to which the $n(n-3)/2$ independent  cross ratios are raised.

\subsubsection{Dilatation Ward identity}

It is straightforward to check that the dimension of the mesh integral \eqref{mesh} equals $\Delta_t - n d$. 
For every edge we have an integration which increases the dimension by $d$ but also a propagator which decreases the dimension by $2\alpha_{ij} + d$.  Each delta functions decreases the dimension by $d$. 
Overall, we then have   
\begin{align}
- \sum_{1 \leq i < j \leq n} 2 \alpha_{ij} - n d = - \sum_{i,j=1}^n \alpha_{ij} - n d = \Delta_t - n d.
\end{align}
The second equality follows from the summation of conditions \eqref{deltaijcond_a} over $m$ from $1$ to $n$.

\subsubsection{Special conformal Ward identities}
\paragraph{Proof by mesh recursion:} 
\label{sec:conf_mesh_P}

To show that the mesh integrals satisfy \eqref{special} we can use their recursive structure as given in  \eqref{meshrec0}. 
First, using the Fourier representation of the Dirac delta function and integration by parts, it is straightforward to show the $1$-point mesh integral satisfies
\[\label{trick2a}
\K^\kappa(0;\,\bs{p}_1) M_1(\bs{p}_1) = 0,
\]
where \eqref{deltaijcond_a} forces $\Delta_1=0$.
Alternatively, the 1-, 2- and 3-point meshes are equivalent to the standard momentum-space 1-, 2- and 3-point functions as shown in section \ref{sec_123ptfns} and hence are conformally invariant. 

We now proceed by induction. We denote by $\mathcal{E}^{(n) \kappa}_{SCWI}$ the action of the special conformal operator in \eqref{special} on the $n$-point mesh integral,
\[\label{meshSCWI}
\mathcal{E}^{(n) \kappa}_{SCWI}(\bs{\Delta}; \bs{p}_1, \ldots, \bs{p}_n) = \K^\kappa(\bs{\Delta}) M_n(\bs{\alpha}; \bs{p}_1,\ldots ,\bs{p}_n).
\]
This expression admits a recursive structure similar to that in \eqref{meshrec0}. 
To see this, let us write the scaling dimensions of the $n$-point function as $\Delta_m^{(n)}$ and those of the $(n-1)$-point function as $\Delta_m^{(n-1)}$.  From \eqref{deltaijcond_a}, these are related by 
\begin{align} \label{deltainc}
\Delta_n^{(n)} = -\sum_{j=1}^{n-1}\alpha_{jn},\qquad
 \Delta_m^{(n)} = \Delta_m^{(n-1)} - \alpha_{mn}, \quad  m = 1, \ldots, n-1.
\end{align}
Thus, given  $n-1$ parameters $\alpha_{mn}$ and a set of $\Delta_m^{(n-1)}$ satisfying \eqref{deltaijcond_a}  at $(n-1)$ points, we can 
construct a solution of \eqref{deltaijcond_a} 
at $n$ points.

Using integration by parts as discussed in appendix \ref{app:conf_mesh_P}, 
one can then show that
\begin{align} \label{recursiveE}
 \mathcal{E}^{(n) \kappa}_{SCWI}(\bs{\Delta}^{(n)}; \bs{p}_1, \ldots, \bs{p}_n )
&=\prod_{i = 1}^{n-1} C_{in} \int \frac{\D^d \bs{q}_{i}}{(2 \pi)^d} \frac{1}{q_1^{2 \alpha_{1n} + d} q_2^{2 \alpha_{2n} + d} \ldots q_{n-1}^{2 \alpha_{n-1,n} + d}}  
\nn\\& \quad
 \times (2 \pi)^d \delta\Big( \bs{p}_n + \sum_{j=1}^{n-1}\bs{q}_j\Big)
\mathcal{E}^{(n-1) \kappa}_{SCWI}(\bs{\Delta}^{(n-1)}; \bs{p}_1, \ldots, \bs{p}_{n-1}),
\end{align}
where the $\bs{q}_i$ are a shorthand for $\bs{q}_{in}$ as in \eqref{meshrec0}.
If the  $(n-1)$-point mesh integral 
satisfies the $(n-1)$-point special conformal Ward identity then  $\mathcal{E}^{(n-1) \kappa}_{SCWI}(\bs{\Delta}^{(n-1)}; \bs{p}_1, \ldots, \bs{p}_{n-1})$ vanishes.  The $n$-point mesh integral then satisfies the $n$-point special conformal Ward identity by virtue of \eqref{recursiveE}.

\paragraph{Proof by total derivatives:} 
\label{sec:conf_mesh_A}

To prove the conformal invariance of the reduced mesh integral $\tilde{M}_n$, one can simply apply the SCWI operator \eqref{special_red} to the integrand of \eqref{mesh_red} and show that the resulting expression is a sum of total derivatives with respect to $\bs{q}_{ij}$. Explicitly, we obtain
\begin{align} \label{to_show_mesh}
& \tilde{\mathcal{K}}^{\kappa}(\bs{\Delta}) \left[ \frac{1}{\text{Den}_n(\bs{\alpha})} \right] = \sum_{\substack{i,j = 1\\i \neq j}}^{n-1} \frac{\partial}{\partial q_{ij}^\mu} \left[ (2 \alpha_{in} + d) \frac{A_{ij}^{\kappa \mu}}{\text{Den}_n(\bs{\alpha})} \right],
\end{align}
where
\begin{align} \label{A_def}
& A_{ij}^{\kappa \mu} = (\delta^{\kappa \mu} \delta_{\alpha \beta} + \delta^\kappa_\beta \delta^\mu_\alpha - \delta^\kappa_\alpha \delta^\mu_\beta ) \frac{q_{ij}^{\alpha} (\bs{l}_i - \bs{p}_i)^\beta}{(\bs{l}_i - \bs{p}_i)^2}.
\end{align}
(Recall that we extended $\q_{ij}$ to $i>j$ by $\bs{q}_{ij} = -\bs{q}_{ji}$ and consequently $\partial/\partial q_{ij}^\mu = - \partial/\partial q_{ji}^\mu$.) Since the right-hand side of \eqref{to_show_mesh} is a sum of total derivatives, it vanishes when the integrals in \eqref{mesh_red} are evaluated. This proves directly the conformal invariance of the mesh integrals. The details of the calculation are presented in appendix \ref{app:conf_mesh_A}.

\subsection{Simplex integrals}\label{simpsubsec}

Having established the conformal invariance of  mesh integrals subject to  \eqref{deltaijcond_a}, we now turn to the general simplex integral \eqref{simplex}. Since the momentum-space cross ratios have vanishing scaling dimension, the dilatation weight of the simplex integral is the same as that of the corresponding mesh integral.  As condition \eqref{deltaijcond_a} implies that the scaling dimension of the simplex integral is $\Delta_t - n d$, the dilatation Ward identity \eqref{dil} is therefore satisfied.

Special conformal invariance can be now established by a number of independent methods. The first method employs a Mellin-Barnes representation of the arbitrary function along with the conformal invariance of the mesh integrals. The second method uses the invariance of mesh integrals to show the action of the special conformal Ward identity on the simplex is implicitly a total derivative.   The third method identifies these total derivatives 
explicitly, generalising the method of total derivatives for mesh integrals in section \ref{sec:conf_mesh_A}.

\subsubsection{Proof by Mellin-Barnes transform} \label{sec:conf_simplex_P}

By proving the  conformal invariance of mesh integrals, we have proven the conformal invariance of simplex integrals where $\hat{f}$ is a monomial in the cross ratios. Indeed, let us consider $\hat{f} = \prod_{I \in \mathcal{U}} \hat{u}_{I}^{\gamma_I}$, where $\gamma_I$ is a set of exponents. For such an $\hat{f}$, on the right-hand side of the simplex integral \eqref{simplex} we find
\begin{align} \label{simplex_to_mesh}
\frac{\prod_{I \in \mathcal{U}} \hat{u}_{I}^{\gamma_I}}{\text{Den}_n(\bs{\alpha})} = \frac{1}{\text{Den}_n(\bs{\alpha}^{(\gamma_1 \ldots \gamma_N)}_{I_1 \ldots I_N})},
\end{align}
where we have $N=n(n-3)/2$ independent cross ratios, $\hat{u}_{I_1}, \ldots, \hat{u}_{I_N}$, and the parameters $\bs{\alpha}^{(\gamma_1 \ldots \gamma_N)}_{I_1 \ldots I_N} = \{ \alpha^{(\gamma_1 \ldots \gamma_N)}_{ij, I_1 \ldots I_N} \}_{1\le i<j\le n}$ are given by
\begin{align} \label{a_abcd}
\alpha^{(\gamma_1 \ldots \gamma_N)}_{ij, I_1 \ldots I_N} = \alpha_{ij} + \sum_{m=1}^N \gamma_m S_{ij, I_m}
\end{align}
where
\begin{align}
S_{ij, [pqrs]} = \delta_{ip} \delta_{jr} + \delta_{iq} \delta_{js} - \delta_{ip} \delta_{jq} - \delta_{ir} \delta_{js}.
\end{align}
The integral now reduces to another mesh integral, but with parameters $\bs{\alpha}^{(\gamma_1 \ldots \gamma_N)}_{I_1 \ldots I_N}$ instead. It is easy to see that the conditions \eqref{deltaijcond_a} are satisfied for the new values of the parameters with the same values of the scaling dimensions. Hence the simplex integral with monomial $\hat{f}$ satisfies the same CWIs as the simplex with $\hat{f} = 1$. This then proves the  conformal invariance of simplex integrals where $\hat{f}$ is given by a multiple Mellin-Barnes transform,
\begin{align}\label{MBfhat}
\hat{f}(\hat{\bs{u}}) = \frac{1}{(2 \pi \I)^N} \int_{c_1 - \I \infty}^{c_1 + \I \infty} \D s_1 \ldots \int_{c_N - \I \infty}^{c_N + \I \infty} \D s_N \: \hat{u}_{I_1}^{s_1} \ldots \hat{u}_{I_N}^{s_N} \hat{F}(s_1, \ldots, s_N).
\end{align}
for some appropriate choice of integration contour specified by $c_1, \ldots c_N$ for which the Mellin-Barnes transform converges.

Finally, let us recall that position-space correlators can themselves be expressed via the Mellin representation \cite{Mack:2009mi, Penedones:2010ue, Fitzpatrick:2011ia}  
\[\label{Mellrep}
\<\O_1(\x_1)\ldots\O_n(\x_n)\>=\frac{1}{(2\pi i)^N}
\int_{c_{ij}-i\infty}^{c_{ij}+i\infty}
[\D \gamma_{ij}] \,\mathcal{M}_n(\gamma_{ij}) \prod_{i<j}^n x_{ij}^{-2\gamma_{ij}}\Gamma(\gamma_{ij}).
\]
Here the integration runs over the $N$ independent solutions of the constraint $\sum_{j\neq i}\gamma_{ij}=\Delta_i$.  
These independent solutions can be parametrised as in \eqref{a_abcd} by writing
\[
-\gamma_{ij}=\alpha_{ij}+\sum_{m=1}^N \gamma_m S_{ij,I_{m}},
\]
where the $\alpha_{ij}$ represent an arbitrary fixed solution of \eqref{deltaijcond_a}.
The Fourier transform of the monomial $\prod x_{ij}^{-2\gamma_{ij}}$ can now be evaluated as described in section \ref{ftmeshsec}.  This yields the simplex representation \eqref{simplex}, where
\[\label{fhatMellin}
\hat{f}(\hat{\bs{u}}) = \frac{1}{(2\pi i)^N}\Big( \prod_{m=1}^N\int_{c_m-i\infty}^{c_m+i\infty}
\D\gamma_{m}\,\hat{u}_{I_m}^{\gamma_m} \Big)\mathcal{M}_n(\gamma_{ij}) \prod_{i<j}^n \pi^{d/2} 2^{d-2\gamma_{ij}}\Gamma(d/2-\gamma_{ij}).
\]
Thus, given the Mellin representation for a position-space correlator, we can immediately write down the Mellin representation for $\hat{f}$.
This formula provides an alternative method of computing $\hat{f}(\hat{\bs{u}})$ for holographic correlators to those we will  discuss in section \ref{holsec}.

\subsubsection{Implicit proof by total derivatives} \label{sec:conf_simplex_K}

Next, we show that the conformal invariance of simplex integrals follows from that of mesh integrals without making use of a Mellin-Barnes representation. In fact, as we will see, it is sufficient to know that simplex integrals with $\hat{f} = 1, \,\hat{u}_I,\, \hat{u}_{I} \hat{u}_{J}$ are conformally invariant. The conformal invariance of these special cases 
follows since they reduce to mesh integrals satisfying the condition \eqref{deltaijcond_a} with the same scaling dimensions.

Since the SCWI operator \eqref{special_red} is a second-order differential operator, its action on the integrand of the simplex integral \eqref{simplex_red} must have the form
\begin{align} \label{toa1}
& \tilde{\mathcal{K}}^{\kappa}(\bs{\Delta}) \left[ \frac{\hat{f}(\hat{\bs{u}})}{\text{Den}_n(\bs{\alpha})} \right] = \hat{f}(\hat{\bs{u}}) C^{\kappa}(\bs{\alpha}) + \frac{\partial \hat{f}(\hat{\bs{u}})}{\partial \hat{u}_I} C^{\kappa}_I(\bs{\alpha}) + \frac{\partial^2 \hat{f}(\hat{\bs{u}})}{\partial \hat{u}_I \partial \hat{u}_J} C^{\kappa}_{IJ}(\bs{\alpha}),
\end{align}
where $C^{\kappa}, \,C^{\kappa}_{I},\, C^{\kappa}_{IJ}$ are coefficients depending on the external and internal momenta as well as the parameters $\bs{\alpha}$, but crucially they are independent of $\hat{f}$.  The indices $I, J$ run over the independent cross ratios and we sum over any repeated indices. 

Conformal invariance for $\hat{f} = 1,\, \hat{u}_I,\, \hat{u}_{I} \hat{u}_{J}$ requires that  the coefficients  $C^{\kappa}, \,C^{\kappa}_{I},\, C^{\kappa}_{IJ}$ are such that the right-hand side of \eqref{toa1} forms a total derivative.  This means they must 
satisfy the conditions
\begin{align}
\hat{f} = 1: \qquad & C^{\kappa} = \sum_{\substack{i,j = 1\\i \neq j}}^{n-1} \frac{\partial}{\partial q_{ij}^\mu} \Gamma_{ij}^{\kappa \mu}\\
\hat{f} =\hat{u}_I : \qquad & C^{\kappa}_{I} = \sum_{\substack{i,j = 1\\i \neq j}}^{n-1} \left( \Gamma_{ij}^{\kappa \mu} \frac{\partial \hat{u}_I}{\partial q_{ij}^\mu} + \frac{\partial}{\partial q_{ij}^\mu} \Gamma^{\kappa \mu}_{ij, I} \right) \\
\hat{f} = \hat{u}_{I} \hat{u}_{J} : \qquad & C^{\kappa}_{IJ}=\sum_{\substack{i,j = 1\\i \neq j}}^{n-1} \Gamma^{\kappa \mu}_{ij, J} \frac{\partial \hat{u}_I}{\partial q_{ij}^\mu}
\end{align}
for some coefficients $\Gamma_{ij}^{\kappa \mu}$ and $\Gamma^{\kappa \mu}_{ij, I}$.  
As these  are independent of $\hat{f}$, however,   it  follows that 
\begin{align} \label{to_show_simp_k}
& \tilde{\mathcal{K}}^{\kappa}(\bs{\Delta}) \left[ \frac{\hat{f}(\hat{\bs{u}})}{\text{Den}_n(\bs{\alpha})} \right] = \sum_{\substack{i,j = 1\\i \neq j}}^{n-1} \frac{\partial}{\partial q_{ij}^\mu} \left[ \Gamma_{ij}^{\kappa \mu}(\bs{\alpha}) \hat{f}(\hat{\bs{u}}) + \sum_{I \in \mathcal{U}} \Gamma^{\kappa \mu}_{ij, I}(\bs{\alpha}) \frac{\partial \hat{f}(\hat{\bs{u}})}{\partial \hat{u}_I}  \right]
\end{align}
for {\it any} $\hat{f}$.  The action of the SCWI operator on the integrand of the simplex is thus a total derivative, and hence the simplex 
is conformally invariant.

\subsubsection{Explicit proof by total derivatives} \label{sec:conf_simplex_A}

The conformal invariance of simplex integrals can also be proven by finding the explicit form of the coefficients $\Gamma_{ij}^{\kappa \mu}$ and $\Gamma^{\kappa \mu}_{ij, I}$ in \eqref{to_show_simp_k} by direct computation. 
This approach is similar to the method of total derivatives for mesh integrals presented in section \ref{sec:conf_mesh_P}.
The result is 
\begin{align}
\Gamma^{\kappa \mu}_{ij}(\bs{\alpha}) & = (2 \alpha_{in} + d) \times \frac{A^{\kappa \mu}_{ij}}{\text{Den}_n(\bs{\alpha})}, \label{G0} \\
\Gamma^{\kappa \mu}_{ij, [pqrs]}(\bs{\alpha}) & = 2 (\delta_{ip} \delta_{rn} + \delta_{iq} \delta_{sn} - \delta_{ip} \delta_{qn} - \delta_{ir} \delta_{sn} ) \times \frac{A^{\kappa \mu}_{ij} \hat{u}_{[pqrs]}}{\text{Den}_n(\bs{\alpha})}, \label{GI}
\end{align}
where the $A_{ij}^{\kappa \mu}$ are defined in \eqref{A_def}.
These formulae can easily be found via  the method presented in the previous subsection, \textit{i.e.}, by substituting $\hat{f} = 1$ and $\hat{f} = \hat{u}_I$. For $\hat{f} = 1$, \eqref{to_show_simp_k} reduces to \eqref{to_show_mesh} and hence $\Gamma^{\kappa \mu}_{ij}$ is given by its right-hand side. For $\hat{f} = \hat{u}_I$, we find 
\begin{equation}\label{Ga}
\Gamma_{ij, I}^{\kappa \mu}(\bs{\alpha}) = \Gamma^{\kappa \mu}_{ij}(\bs{\alpha}_{I}) - \Gamma^{\kappa \mu}_{ij}(\bs{\alpha}) \hat{u}_{I},
\end{equation}
where $\bs{\alpha}_{I} = \bs{\alpha}_{I}^{(1)}$ is given by \eqref{a_abcd} with $\gamma_1 = 1$. Using the explicit form of $\Gamma^{\kappa \mu}_{ij}$, this expression then reduces to \eqref{GI}.

Equation \eqref{to_show_simp_k} is guaranteed to hold by the indirect argument in the previous subsection, we can also show it directly through a straightforward (but cumbersome) calculation.  We present this calculation in appendix \ref{app:conf_simp_A}.

\subsubsection{Generality  of the simplex representation}

Having established that simplex integrals solve the conformal Ward identities, a natural question to ask is whether {\it all} solutions can be represented in this form.
Since simplex integrals involve an arbitrary function $\hat{f}(\hat{\bs{u}})$ of the momentum-space cross ratios, the  dimension of the solution space is the same as that in position space, where we have an arbitrary function $f(\bs{u})$ of the same number of cross ratios.
Moreover, whenever the position-space solution admits a Mellin representation \eqref{Mellrep}, a corresponding simplex solution exists with $\hat{f}(\hat{\bs{u}})$ given by \eqref{fhatMellin}.  
More generally, the proofs of conformal invariance above require only that  $\hat{f}(\hat{\bs{u}})$ is twice differentiable, and that the simplex integral converges.\footnote{When the integral diverges the correlators can still be understood in terms of renormalisation following \cite{Bzowski:2015pba, Bzowski:2017poo,Bzowski:2018fql}, see \cite{Bzowski:2019kwd}.  These cases are associated with additional local solutions of the conformal Ward identities, and the renormalised correlators satisfy anomalous Ward identities. Understanding the renormalisation of general conformal $n$-point functions in position space is an open problem.} 
For these reasons, we expect that the simplex representation  spans at least a dense subset of the possible solutions of the momentum-space conformal Ward identities. Making this precise, however, and proving completeness in full generality 
requires specifying in sufficient detail the function space we are working on, and this is outside the scope of this work.

\section{Simplex representation for holographic correlators}
\label{holsec}

In this section, we explore how correlators for holographic CFTs can be written in simplex form.
Since exchange diagrams can be decomposed into a sum of contact diagrams \cite{DHoker:1999mqo,DHoker:1999kzh}, we will restrict our attention here to contact diagrams leaving exchanges to future work.

Our goal is thus to find the specific form of the function $\hat{f}(\hat{\bs{u}})$ of momentum-space ratios  entering the simplex representation of the $n$-point contact diagram.
As a warm-up, we begin with the case of the $4$-point function following an approach inspired by the star-mesh duality from electrical circuit theory.  This allows us to map the $4$-point contact diagram to a corresponding tetrahedral diagram from which we can read off the form of $\hat{f}(\hat{\bs{u}})$.
To deal with the general $n$-point function, we then discuss (for variety) an alternative approach starting from position space followed by a recursive application of the convolution theorem.

\subsection{Star-mesh duality for Witten diagrams}

\begin{figure}
\centering
\hspace{2mm}\begin{tikzpicture}[scale=3.65]
\def \start{0.106066};
\def \del{0.0176777};
\def \var{0.035355};
\def \disp{-1.9};

\draw[black,fill=black] (0,0) circle [radius=0.02];
\draw[black,fill=black] (1,0) circle [radius=0.02];
\draw[black,fill=black] (1/2,1.1) circle [radius=0.02];

\draw (0,0) -- (0.35,0);
\draw (0.65,0) -- (1,0);
\draw (0,0) -- (0.144831,0.318628);
\draw (0.268972,0.591738) -- (0.5,1.1);
\draw (1,0) --  (1-0.144831,0.318628);
\draw  (1-0.268972,0.591738) -- (1-0.5,1.1);

\draw (0.144831,0.318628) -- (0.109658,0.362078) --(0.221385,0.366216) --
(0.151038,0.453114) -- (0.262765,0.457252)--(0.192418,0.544151)--
(0.304145,0.548289)--(0.268972,0.591738);

\draw (1-0.144831,0.318628) -- (1-0.109658,0.362078) --(1-0.221385,0.366216) --
(1-0.151038,0.453114) -- (1-0.262765,0.457252)--(1-0.192418,0.544151)--
(1-0.304145,0.548289)--(1-0.268972,0.591738);

\draw (0.35,0) -- (0.375,0.05) -- (0.05+0.375,-0.05) -- (0.05+0.425,0.05) 
-- (0.05+0.475,-0.05) -- (0.05+0.525,0.05) -- (0.05+0.575,-0.05) -- (0.05+0.6,0);

\node[right] at (0.37,-0.15) {$z_{23}$};
\node[right] at (-0.15,0.5) {$z_{13}$};
\node[right] at (0.55+0.35,0.5) {$z_{12}$};

\node at (-0.5,0.5) {$\equiv$};


\draw[black,fill=black] (0+\disp,0) circle [radius=0.02];
\draw[black,fill=black] (1+\disp,0) circle [radius=0.02];
\draw[black,fill=black] (1/2+\disp,1.1) circle [radius=0.02];

\draw (\disp+0,0) -- (\disp+0.3-\start,0.3-\start);
\draw (\disp+0.3+\start,0.3+\start) -- (\disp+0.5,0.5); 
\draw (\disp+0.5,0.5) -- (\disp+0.7-\start,0.3+\start);
\draw (\disp+0.7+\start,0.3-\start) -- (\disp+1,0);
\draw (\disp+0.5,0.5) -- (\disp+0.5,0.65);
\draw (\disp+0.5,0.95) -- (\disp+0.5,1.1);

\draw (\disp+0.5,0.65) -- (\disp+0.5-0.05,0.675) -- (\disp+0.5+0.05,0.725) -- (\disp+0.5-0.05,0.775) -- (\disp+0.5+0.05,0.825) -- (\disp+0.5-0.05,0.875) -- (\disp+0.5+0.05,0.925) -- (\disp+0.5,0.95);

\draw (\disp+0.3-\start,0.3-\start) -- (\disp+0.3-\start+\del-\var,0.3-\start+\del+\var) -- (\disp+0.3-\start+3*\del+\var,0.3-\start+3*\del-\var) -- (\disp+0.3-\start+5*\del-\var,0.3-\start+5*\del+\var) -- (\disp+0.3-\start+7*\del+\var,0.3-\start+7*\del-\var) -- (\disp+0.3-\start+9*\del-\var,0.3-\start+9*\del+\var) -- (\disp+0.3-\start+11*\del+\var,0.3-\start+11*\del-\var) -- (\disp+0.3+\start,0.3+\start);
\draw (\disp+0.7-\start,0.3+\start) 
  -- (\disp+0.7-\start+\del+\var, 0.3+\start-\del+\var) 
  -- (\disp+0.7-\start+3*\del-\var, 0.3+\start-3*\del-\var) 
  -- (\disp+0.7-\start+5*\del+\var, 0.3+\start-5*\del+\var) 
  -- (\disp+0.7-\start+7*\del-\var, 0.3+\start-7*\del-\var) 
  -- (\disp+0.7-\start+9*\del+\var, 0.3+\start-9*\del+\var) 
  -- (\disp+0.7-\start+11*\del-\var, 0.3+\start-11*\del-\var) 
  -- (\disp+0.7+\start, 0.3-\start);

\node[left] at (\disp+0.25,0.375) {$Z_3$};
\node[right] at (\disp+0.75,0.375) {$Z_2$};
\node[right] at (\disp+0.57,0.8) {$Z_1$};
\end{tikzpicture}

\vspace{3mm}

\begin{tikzpicture}[scale=3.65]
\def \start{0.106066};
\def \del{0.0176777};
\def \var{0.035355};
\def \disp{-1.9};

\draw (0.35,0) -- (0,0) -- (0,0.35);
\draw (0.65,0) -- (1,0) -- (1,0.35);
\draw (0.35,1) -- (0,1) -- (0,0.65);
\draw (0.65,1) -- (1,1) -- (1,0.65);

\draw[black,fill=black] (0,0) circle [radius=0.02];
\draw[black,fill=black] (1,0) circle [radius=0.02];
\draw[black,fill=black] (0,1) circle [radius=0.02];
\draw[black,fill=black] (1,1) circle [radius=0.02];

\draw (0,0.35) -- (-0.05,0.375) -- (0.05,0.425) -- (-0.05,0.475) -- (0.05,0.525) -- (-0.05,0.575) -- (0.05,0.625) -- (0,0.65);
\draw (1,0.35) -- (1-0.05,0.375) -- (1.05,0.425) -- (1-0.05,0.475) -- (1.05,0.525) -- (1-0.05,0.575) -- (1.05,0.625) -- (1,0.65);
\draw (0.35,0) -- (0.375,0.05) -- (0.425,-0.05) -- (0.475,0.05) -- (0.525,-0.05) -- (0.575,0.05) -- (0.625,-0.05) -- (0.65,0);
\draw (0.35,1) -- (0.375,1.05) -- (0.425,1-0.05) -- (0.475,1.05) -- (0.525,1-0.05) -- (0.575,1.05) -- (0.625,1-0.05) -- (0.65,1);

\draw (0,1) -- (0.3-\start,0.7+\start);
\draw (1,0) -- (0.3+\start,0.7-\start);
\draw (0.3-\start,0.7+\start) 
  -- (0.3-\start+\del+\var, 0.7+\start-\del+\var) 
  -- (0.3-\start+3*\del-\var, 0.7+\start-3*\del-\var) 
  -- (0.3-\start+5*\del+\var, 0.7+\start-5*\del+\var) 
  -- (0.3-\start+7*\del-\var, 0.7+\start-7*\del-\var) 
  -- (0.3-\start+9*\del+\var, 0.7+\start-9*\del+\var) 
  -- (0.3-\start+11*\del-\var, 0.7+\start-11*\del-\var) 
  -- (0.3+\start, 0.7-\start);

\draw (0,0) -- (0.3-\start,0.3-\start);
\draw (0.3+\start,0.3+\start) -- (0.5-0.04,0.5-0.04);
\draw (0.5+0.04,0.5+0.04) -- (1,1);
\draw (0.3-\start,0.3-\start) -- (0.3-\start+\del-\var,0.3-\start+\del+\var) -- (0.3-\start+3*\del+\var,0.3-\start+3*\del-\var) -- (0.3-\start+5*\del-\var,0.3-\start+5*\del+\var) -- (0.3-\start+7*\del+\var,0.3-\start+7*\del-\var) -- (0.3-\start+9*\del-\var,0.3-\start+9*\del+\var) -- 
(0.3-\start+11*\del+\var,0.3-\start+11*\del-\var) -- (0.3+\start,0.3+\start);

\node[right] at (0.05,0.50) {$z_{34}$};
\node[left] at (0.95,0.50) {$z_{12}$};
\node at (0.5,1.125) {$z_{14}$};
\node at (0.5,-0.15) {$z_{23}$};
\node[right] at (0.325,0.75) {$z_{24}$};
\node[right] at (0.30,0.20) {$z_{13}$};

\node at (-0.5,0.5) {$\equiv$};

\draw[black,fill=black] (0+\disp,0) circle [radius=0.02];
\draw[black,fill=black] (1+\disp,0) circle [radius=0.02];
\draw[black,fill=black] (0+\disp,1) circle [radius=0.02];
\draw[black,fill=black] (1+\disp,1) circle [radius=0.02];

\draw (\disp+0,0) -- (\disp+0.3-\start,0.3-\start);
\draw (\disp+0.3+\start,0.3+\start) -- (\disp+0.7-\start,0.7-\start);
\draw (\disp+0.7+\start,0.7+\start) -- (\disp+1,1);
\draw (\disp+0,1) -- (\disp+0.3-\start,0.7+\start);
\draw (\disp+0.3+\start,0.7-\start) -- (\disp+0.7-\start,0.3+\start);
\draw (\disp+0.7+\start,0.3-\start) -- (\disp+1,0);

\draw (\disp+0.3-\start,0.3-\start) -- (\disp+0.3-\start+\del-\var,0.3-\start+\del+\var) -- (\disp+0.3-\start+3*\del+\var,0.3-\start+3*\del-\var) -- (\disp+0.3-\start+5*\del-\var,0.3-\start+5*\del+\var) -- (\disp+0.3-\start+7*\del+\var,0.3-\start+7*\del-\var) -- (\disp+0.3-\start+9*\del-\var,0.3-\start+9*\del+\var) -- (\disp+0.3-\start+11*\del+\var,0.3-\start+11*\del-\var) -- (\disp+0.3+\start,0.3+\start);
\draw (\disp+0.7-\start,0.7-\start) -- (\disp+0.7-\start+\del-\var,0.7-\start+\del+\var) -- (\disp+0.7-\start+3*\del+\var,0.7-\start+3*\del-\var) -- (\disp+0.7-\start+5*\del-\var,0.7-\start+5*\del+\var) -- (\disp+0.7-\start+7*\del+\var,0.7-\start+7*\del-\var) -- (\disp+0.7-\start+9*\del-\var,0.7-\start+9*\del+\var) -- (\disp+0.7-\start+11*\del+\var,0.7-\start+11*\del-\var) -- (\disp+0.7+\start,0.7+\start);
\draw (\disp+0.3-\start,0.7+\start) 
  -- (\disp+0.3-\start+\del+\var, 0.7+\start-\del+\var) 
  -- (\disp+0.3-\start+3*\del-\var, 0.7+\start-3*\del-\var) 
  -- (\disp+0.3-\start+5*\del+\var, 0.7+\start-5*\del+\var) 
  -- (\disp+0.3-\start+7*\del-\var, 0.7+\start-7*\del-\var) 
  -- (\disp+0.3-\start+9*\del+\var, 0.7+\start-9*\del+\var) 
  -- (\disp+0.3-\start+11*\del-\var, 0.7+\start-11*\del-\var) 
  -- (\disp+0.3+\start, 0.7-\start);
\draw (\disp+0.7-\start,0.3+\start) 
  -- (\disp+0.7-\start+\del+\var, 0.3+\start-\del+\var) 
  -- (\disp+0.7-\start+3*\del-\var, 0.3+\start-3*\del-\var) 
  -- (\disp+0.7-\start+5*\del+\var, 0.3+\start-5*\del+\var) 
  -- (\disp+0.7-\start+7*\del-\var, 0.3+\start-7*\del-\var) 
  -- (\disp+0.7-\start+9*\del+\var, 0.3+\start-9*\del+\var) 
  -- (\disp+0.7-\start+11*\del-\var, 0.3+\start-11*\del-\var) 
  -- (\disp+0.7+\start, 0.3-\start);

\node[left] at (\disp+0.25,0.375) {$Z_3$};
\node[left] at (\disp+0.25,1-0.375) {$Z_4$};
\node[right] at (\disp+0.75,0.375) {$Z_2$};
\node[right] at (\disp+0.75,1-0.375) {$Z_1$};
\end{tikzpicture}
\caption{Equivalent electrical networks of resistors under star-mesh duality, where the conductivities and currents are related as given in \eqref{starmesh1}.   
The external currents flowing into the corresponding dotted nodes and the overall power dissipation are equal. 
 \label{fig:4}}
\end{figure}
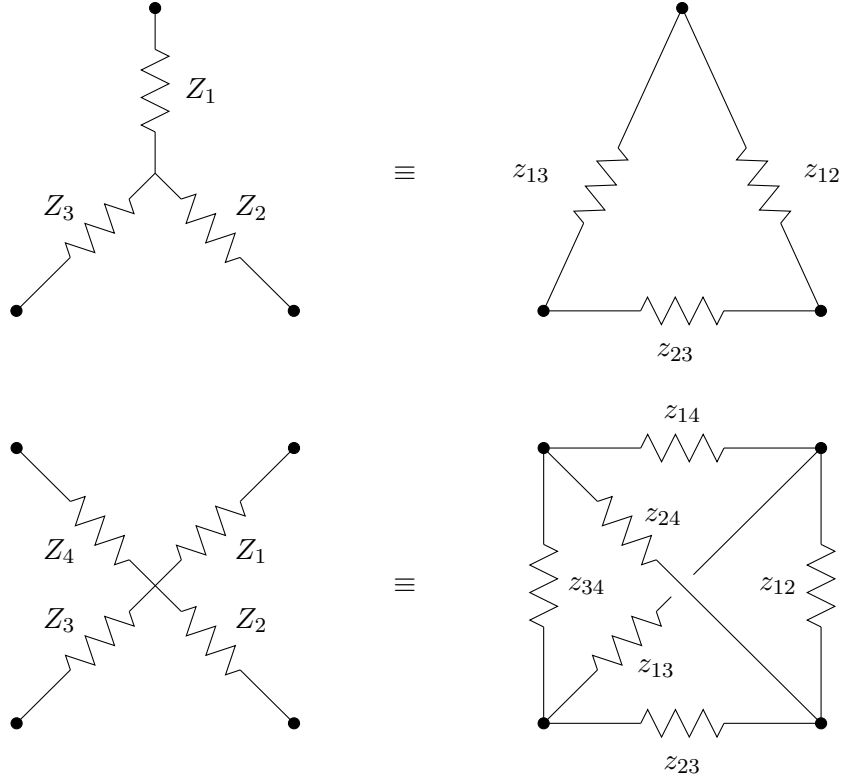

In momentum space, the $n$-point contact diagram consists of $n$ bulk-to-boundary propagators interacting at a common bulk point with radial coordinate $z$ over which we integrate.   Each propagator is constructed from a modified Bessel function, and with the standard holographic normalisation we find
\begin{align}\label{nptcontact}
\mathcal{I}_n &\equiv \lla \O_1(\p_1)\ldots\O_n(\p_n)\rra_{contact} =
\int_0^\infty \frac{\D z}{z^{d+1}}\prod_{j=1}^n \frac{2^{1-\beta_j}}{\Gamma(\beta_j)} z^{d/2}p_j^{\beta_j}K_{\beta_j}(p_j z),
\end{align}
where $\beta_j=\Delta_j-d/2$. 
Schwinger parametrising the Bessel functions as
\[\label{SchwingerK}
p_j^{\beta_j}K_{\beta_j}(p_j z) = \frac{1}{2}z^{\beta_j}\int_0^\infty\D Z_j \,Z_j^{\beta_j-1}\exp\Big[-\frac{1}{2}\Big(\frac{p_j^2}{Z_j}+z^2 Z_j\Big)\Big]
\]
and doing the $z$ integral we then find
\begin{align}
\mathcal{I}_n
=  \hat{C}_n \,\Big(\prod_{j=1}^n\int_{0}^\infty \D Z_j \,Z_j^{\beta_j -1} \Big)Z_t^{(d-\Delta_t)/2}\exp\Big(-\sum_{j=1}^n\frac{p_j^2}{2 Z_j}\Big)  \label{form1}
\end{align}
where   
\[
\hat{C}_n = 2^{(n-1)d/2-\Delta_t/2-1}\Gamma\Big(\frac{\Delta_t-d}{2}\Big)\prod_{j=1}^n\frac{1}{\Gamma(\beta_j)}, \qquad
Z_t = \sum_{j=1}^n Z_j.
\]
If we regard the Schwinger parameters $Z_j$ as conductivities, and the momenta $\p_j$ as ingoing currents, then the exponent in \eqref{form1} describes the power dissipation in the star-shaped electrical network, as shown (for $n=3,4$) in figure \ref{fig:4}.  
A well-known result from electrical circuit theory (see {\it e.g.,} \cite{starmeshpaper0, starmeshpaper1, starmeshpaper2})  states that this $n$-star network is then equivalent to a corresponding $(n-1)$-simplex or `mesh' network.  For the $4$-point function, this has the form of a tetrahedron as illustrated.
The current ({\it i.e.,} momentum) $\bs{i}_{jk}$ flowing from vertex $j$ to $k$ of the simplex, and the conductivity ({\it i.e.,} Schwinger parameter) $z_{jk}$ of each leg, are both fixed in terms of the currents and conductivities of the original star network. 
Of course, for genuine electrical circuits currents 
 are purely scalar quantities, but as we now show there is a straightforward vectorial generalisation.

This is the star-mesh relation 
\[
z_{jk}=\frac{Z_j Z_k}{Z_t}, \qquad \bs{i}_{jk} = \frac{1}{Z_t}(\p_j Z_k - \p_k Z_j),  \label{starmesh1}
\]
where our convention for currents is that $\bs{i}_{jk}=-\bs{i}_{kj}$ and $\bs{i}_{jj}$ (with no sum) vanishes.
We can immediately verify momentum conservation at each vertex of the simplex, analogous to current conservation or Kirchoff's first law:
\[
\sum_{k} \bs{i}_{jk} =\frac{1}{Z_t} \Big(\p_j\sum_{k}Z_k
-Z_j\sum_{k}\p_k\Big) 
= \p_j.
\]
In the second equality here, we used conservation of the external momenta.

Kirchoff's second law, the vanishing of the voltage drop around every closed loop, now becomes the vectorial relation
\[\label{voltdrop}
0=\frac{\bs{i}_{jk}}{z_{jk}}+\frac{\bs{i}_{kl}}{z_{kl}}+\frac{\bs{i}_{lj}}{z_{lj}} \qquad \forall \, j,k,l,
\]
where the  `voltage drop' from vertex $j$ to $k$ is $\bs{i}_{jk}/z_{jk} = \p_j/Z_j - \p_k/Z_k$.

The power dissipated in both networks is moreover precisely the same,
\[\label{samepower}
\sum_{j} \frac{p_j^2}{Z_j} = \sum_{j<k}\frac{i_{jk}^2}{z_{jk}},
\]
as can be verified by straightforward manipulation:
\begin{align}
\sum_{j<k}\frac{i_{jk}^2}{z_{jk}} &= \frac{1}{Z_t}\sum_{j<k}\Big(\frac{Z_k}{Z_j}p_j^2+\frac{Z_j}{Z_k}p_k^2-2\p_j\cdot\p_k\Big)
\nn\\&
=\frac{1}{Z_t}\Big(\sum_{j} \frac{1}{Z_j}p_j^2\Big(\sum_{k\neq j}Z_k\Big) - \sum_j \p_j\cdot\Big(\sum_{k\neq j}\p_k\Big)\Big)\nn\\
&= \frac{1}{Z_t}\sum_j\Big( \frac{Z_t-Z_j}{Z_j}p_j^2 +p_j^2\Big)
=\sum_{j} \frac{p_j^2}{Z_j}. 
\end{align}
The currents also satisfy the interesting relation
\[\label{Kleinquadric}
0 = \bs{i}_{j[k}\cdot \bs{i}_{lm]} =  \bs{i}_{jk}\cdot \bs{i}_{lm}+\bs{i}_{jm}\cdot \bs{i}_{kl}+\bs{i}_{jl}\cdot \bs{i}_{mk} \qquad \forall \, j,k,l,m.
\]
This suggests a possible connection to the Klein correspondence,  which we will return to in the conclusions.

\subsubsection{3-point function}\label{startrisection}

As a warm-up, let us begin with the $3$-point function before moving on to tackle the $4$-point function.   In \cite{Bzowski:2013sza}, the triple-$K$ integral $\mathcal{I}_3$ was shown to be equivalent to a 1-loop triangle Feynman integral. Here, we revisit this result from the perspective of star-mesh duality.  The calculation has two steps.  First, we re-express the star form of the integral \eqref{form1} in terms of the conductivities and currents of the corresponding mesh, in this case a triangle as shown in figure \ref{fig:4}.  
Since the conductivities correspond to Schwinger parameters, this change of variables introduces a corresponding Jacobian factor.
Second, we rewrite this Jacobian (and a similar factor present in the original integral) in the form of a loop integral.  To do this, we  introduce an internal current running around the triangle: indeed, we know such internal currents must be present from the form of the simplex representation.
Integrating out the Schwinger parameters then leaves us with a pure triangle Feynman integral.

To convert $\mathcal{I}_3$ in \eqref{form1} to triangle form, we need to invert \eqref{starmesh1} to express the star conductivities $Z_j$ in terms of those of the triangle, $z_{jk}$.  Since there are three of each, this mapping is one-to-one:
\[\label{startriZ}
Z_1 = \frac{\mu}{z_{23}}, \qquad Z_2 = \frac{\mu}{z_{13}},\qquad
Z_3 = \frac{\mu}{z_{12}},\qquad \mu = z_{12}z_{23}+z_{23}z_{13}+z_{13}z_{12}.
\]
Making use of \eqref{samepower}, and the Jacobian 
\[
\prod_{j=1}^3 \D Z_j = \frac{\mu^3}{(z_{12}z_{23}z_{13})^2} \,\D z_{12}\,\D z_{23}\,\D z_{13},
\]
we then find
\[\label{i3res1}
\mathcal{I}_3 = \hat{C}_3 \,\Big(\prod_{j<k}\int_0^\infty\D z_{jk}\, z_{jk}^{\Delta_j+\Delta_k-\Delta_t/2-1}\Big)
 \mu^{-d/2} \exp\Big(-\sum_{j<k}\frac{i_{jk}^2}{2z_{jk}}\Big).
\]
Introducing an internal loop current $\bs{j}$, we now define
\[
\bs{i}'_{12} = \bs{i}_{12} + \bs{j}, \qquad \bs{i}'_{23} = \bs{i}_{23} + \bs{j}, \qquad \bs{i}'_{13} = \bs{i}_{13} - \bs{j}, 
\]
leaving all the external currents the same,
\[
\sum_k \bs{i}'_{jk} =\sum_k \bs{i}_{jk} = \p_j.
\]
Integrating out this internal current yields
\begin{align}
\int\D^d \bs{j} \,\exp \Big(-\sum_{j<k}\frac{{i'}_{jk}^2}{2z_{jk}}\Big)= \Big(\frac{2\pi z_{12}z_{23}z_{13}}{\mu}\Big)^{d/2} \exp\Big(-\sum_{j<k}\frac{i_{jk}^2}{2z_{jk}}\Big)
\end{align}
since as a consequence of the vanishing of the voltage drop around closed loops \eqref{voltdrop}, all $\bs{j}\cdot\bs{i}_{jk}$ cross-terms in the expansion of the exponent vanish:
\begin{align}\label{nocrossterms}
\sum_{j<k}\frac{{i'}_{jk}^2}{2z_{jk}} &= \frac{\mu }{2z_{12}z_{23}z_{13}} j^2+ \bs{j}\cdot\Big(\frac{\bs{i}_{12}}{z_{12}}+\frac{\bs{i}_{23}}{z_{23}}-\frac{\bs{i}_{13}}{z_{13}}\Big) + \sum_{j<k}\frac{i_{jk}^2}{2z_{jk}} 
= \frac{\mu}{2z_{12}z_{23}z_{13}}j^2 + \sum_{j<k}\frac{i_{jk}^2}{2z_{jk}}.
\end{align}
We can thus exchange the factor of $\mu^{-d/2}$ in \eqref{i3res1} for an integral over the internal current:
\begin{align}
\mathcal{I}_3 = (2\pi)^{-d/2}\hat{C}_3 \,\int\D^d\bs{j}\, \Big(\prod_{j<k}\int_0^\infty\D z_{jk}\, z_{jk}^{\Delta_j+\Delta_k-\Delta_t/2-d/2-1}\Big)
  \exp\Big(-\sum_{j<k}\frac{{i'}_{jk}^2}{2z_{jk}}\Big).
\end{align}
In fact, we can simplify further by changing to a shifted integration variable 
\[
\bs{q} = \bs{j} +\bs{i}_{12},
\]
meaning 
\[
\bs{i}'_{12} = \bs{q}, \qquad \bs{i}'_{23}=\p_2+\q,\qquad \bs{i}'_{13}=\p_1-\q.
\]
It is now straightforward to perform the remaining integrations over the $z_{jk}$.  This yields  the 1-loop triangle, or simplex representation of the triple-$K$ integral:
\begin{align}\label{I3tri}
\mathcal{I}_3 = \tilde{C}_3\,\int\frac{\D^d\bs{q}}{(2\pi)^d}\,\frac{1}{|\q|^{2\alpha_{12}+d}|\q-\p_1|^{2\alpha_{13}+d}|\q+\p_2|^{2\alpha_{23}+d}}
\end{align}
where
\[
\tilde{C}_3 = 2^{2d-\Delta_t/2}\pi^{d/2}\hat{C}_3 \prod_{j<k} \Gamma(\alpha_{jk}+d/2), \qquad 
\alpha_{jk} = -\Delta_j-\Delta_k+\frac{\Delta_t}{2}.
\]
This result agrees with \eqref{triint}, as it must since the 3-point function is unique.  Comparing with in appendix A.3 of \cite{Bzowski:2013sza}, we recognise the change of Schwinger parameters there (equation (A.3.12)) as the star-triangle mapping  \eqref{startriZ}.

\subsubsection{4-point function}

Let us now return to the $4$-point contact diagram.
The general procedure mirrors that for the $3$-point function above.  First, we convert the star form of $\mathcal{I}_4$ in \eqref{form1} to the corresponding mesh form.  This entails replacing all the star conductivities $Z_j$ with their mesh equivalents $z_{jk}$, and replacing the exponent using \eqref{samepower}, the equality of the power dissipation.  Next, we introduce internal loop currents running around all the faces of the tetrahedron, since we know these must be present in the simplex representation.  Integrating over these currents then 
simplifies the result in two ways: firstly, by removing a certain prefactor involving the $z_{jk}$, and secondly, by permitting a shift of the currents.  The result then takes the desired simplex form \eqref{kint4}.

The main difference for $n\ge 4$ is that the mesh network now has more resistors than the corresponding star network, so the mapping of resistors is no longer one-to-one.
For the tetrahedron, we have six resistors in total though clearly only four of these are independent, since from \eqref{starmesh1} all cross ratios of the $z_{jk}$ are unity. 
To eliminate this freedom, we can choose a parametrisation centred around a particular vertex of the tetrahedron.  Taking this as the fourth vertex, we select as independent variables the set $z_{14}$, $z_{24}$, $z_{34}$ and 
\begin{align}
\lambda=z_{12}z_{34}=z_{13}z_{24}=z_{14}z_{23}.
\end{align}
The remaining conductivities are then
\[
z_{12}=\frac{\lambda}{z_{34}}, \qquad z_{13}=\frac{\lambda}{z_{24}}, \qquad
z_{23}=\frac{\lambda}{z_{14}}, 
\]
while the conductivities of the original $4$-star network are
\begin{align}
 Z_4 = \frac{\rho}{\lambda}, \qquad 
Z_i &= \frac{\rho \,z_{i4}}{z_{14}z_{24}z_{34}}, \qquad i=1,2,3,
\end{align}
where
\[
\rho = z_{14}z_{24}z_{34}+\lambda(z_{14}+z_{24}+z_{34}). 
\]
Evaluating the Jacobian
\[
\prod_{j=1}^4 \D Z_j= \frac{\rho^4}{\lambda^2 (z_{14}z_{24}z_{34})^3} \,\D\lambda \prod_{i=1}^3 \D z_{i4},
\]
the star form of the $4$-point contact diagram \eqref{form1} can now be rewritten in the corresponding tetrahedral form
\begin{align}\label{I4_intres}
\mathcal{I}_4
&= \hat{C}_4\,\int_0^\infty \D\lambda \, \lambda^{\delta}\Big(\prod_{i=1}^3\int_0^\infty \D z_{i4} \,z_{i4}^{\delta_i}\Big) \rho^{-d} \nn\\&\qquad \times \exp\Big[-\frac{1}{2}\Big(\frac{i_{14}^2}{z_{14}}+\frac{i_{24}^2}{z_{24}}+\frac{i_{34}^2}{z_{34}}+\frac{z_{14}}{\lambda} i_{23}^2+\frac{z_{24}}{\lambda} i_{13}^2+\frac{z_{34}}{\lambda} i_{12}^2\Big)\Big]
\end{align}
where we replaced the exponent using \eqref{samepower} and defined for convenience
\begin{align}
\delta = \frac{\Delta_t}{2}-\Delta_4-1, \qquad
\delta_i =\Delta_i+\Delta_4 -\frac{\Delta_t}{2}+\frac{d}{2}-1.
\end{align}

Let us now introduce a set of internal currents ({\it i.e.,} momenta) running around the faces of the tetrahedron.
Numbering these currents according to the vertex opposite, and choosing a circulation given by the right-hand rule oriented towards this vertex, we define
\begin{align}\label{iprimedef}
\bs{i'}_{12}&=\bs{i}_{12}-\bs{j}_3+\bs{j}_4,\qquad
\bs{i'}_{23}=\bs{i}_{23}-\bs{j}_1+\bs{j}_4,\qquad
\bs{i'}_{13}=\bs{i}_{13}+\bs{j}_2-\bs{j}_4\nn,\\
\bs{i'}_{14}&=\bs{i}_{14}-\bs{j}_2+\bs{j}_3,\qquad
\bs{i'}_{24}=\bs{i}_{24}+\bs{j}_1-\bs{j}_3,\qquad
\bs{i'}_{34}=\bs{i}_{34}-\bs{j}_1+\bs{j}_2.
\end{align}
As these currents are purely internal, all the external currents are unchanged:
\[
\sum_{k}\bs{i'}_{jk} =  \sum_{k}\bs{i}_{jk}=
\bs{p}_j. 
\]
Integrating out these internal currents, we find
\begin{align}
&\int\D^d\bs{j}_1\D^d\bs{j}_2\D^d\bs{j}_3 
 \exp\Big[-\frac{1}{2}\Big(\frac{{i'}_{14}^2}{z_{14}}+\frac{{i'}_{24}^2}{z_{24}}+\frac{{i'}_{34}^2}{z_{34}}+\frac{z_{14}}{\lambda} {i'}_{23}^2+\frac{z_{24}}{\lambda} {i'}_{13}^2+\frac{z_{34}}{\lambda} {i'}_{12}^2\Big)\Big] \nn\\[1ex]&
 =\Big(\frac{8 \pi^3 z_{14}z_{24}z_{34}\lambda^3}{\rho^2}\Big)^{d/2} \exp\Big[-\frac{1}{2}\Big(\frac{i_{14}^2}{z_{14}}+\frac{i_{24}^2}{z_{24}}+\frac{i_{34}^2}{z_{34}}+\frac{z_{14}}{\lambda} i_{23}^2+\frac{z_{24}}{\lambda} i_{13}^2+\frac{z_{34}}{\lambda} i_{12}^2\Big)\Big]
\end{align}
where the dependence on $\bs{j}_4$ cancels out.  
Just as we saw in \eqref{nocrossterms}, all the $\bs{j}_k\cdot\bs{i}_{lm}$ cross-terms cancel when the exponent of the left-hand side is expanded out.
This is because each $\bs{j}_k$ is dotted, after collecting terms, with the sum of the `voltage drop' around a closed loop, which vanishes according to \eqref{voltdrop}.  The three Gaussian integrals over the $\bs{j}_k$ then generate the prefactor shown.

We can thus replace the factor of $\rho^{-d}$ in \eqref{I4_intres} by an integration over internal currents:
\begin{align}
\mathcal{I}_4
&=(2\pi)^{-3d/2} \hat{C}_4\,\int_0^\infty \D\lambda \, \lambda^{\delta-3d/2}\Big(\prod_{i=1}^3\int\D^d\bs{j}_i \int_0^\infty \D z_{i4} \,z_{i4}^{\delta_i-d/2}\Big) \nn\\&\qquad \times \exp\Big[-\frac{1}{2}\Big(\frac{{i'}_{14}^2}{z_{14}}+\frac{{i'}_{24}^2}{z_{24}}+\frac{{i'}_{34}^2}{z_{34}}+\frac{z_{14}}{\lambda} {i'}_{23}^2+\frac{z_{24}}{\lambda} {i'}_{13}^2+\frac{z_{34}}{\lambda} {i'}_{12}^2\Big)\Big]. 
\end{align}
Exchanging the $\bs{j}_k$ for the shifted currents
\begin{align}\label{shift1}
\bs{q}_1&=\bs{i}_{23}-\bs{j}_1 +\bs{j}_4,\\
\bs{q}_2&=-\bs{i}_{13}-\bs{j}_2+\bs{j}_4,\\
\bs{q}_3&=\bs{i}_{12}-\bs{j}_3+\bs{j}_4,\label{shift3}
\end{align}
then allows us to simplify further to 
\begin{align}
\mathcal{I}_4
&=(2\pi)^{-3d/2} \hat{C}_4\,\int_0^\infty \D \lambda \, \lambda^{\delta-3d/2}\Big(\prod_{i=1}^3\int\D^d\bs{q}_i \int_0^\infty \D z_{i4} \,z_{i4}^{\delta_i-d/2}\Big) \nn\\&\qquad \times \exp\Big[-\frac{1}{2}\Big(\frac{1}{z_{14}}|\p_1+\q_2-\q_3|^2+\frac{1}{z_{24}}|\p_2+\q_3-\q_1|^2+\frac{1}{z_{34}}|\p_3+\q_1-\q_2|^2\nn\\[1ex]& \qquad\qquad\qquad\qquad  +\frac{z_{14}}{\lambda} q_1^2+\frac{z_{24}}{\lambda} q_2^2+\frac{z_{34}}{\lambda} q_3^2\Big)\Big].
\end{align}
To reach the desired simplex representation, we now have two tasks remaining.  First, we need to generate the necessary denominator \eqref{Den3}, and second, we need to to verify that the rest of the integral depends only on the momentum-space cross ratios in  \eqref{uvhatdef}.  

Both tasks can be accomplished simultaneously by rescaling $z_{24}\rightarrow z_{24}|\p_2+\q_3-\q_1|^2$ and replacing $\lambda = q_2^2 |\p_2+\q_3-\q_1|^2/z^{2}$,  then performing the $z_{i4}$ integrals using \eqref{SchwingerK}.   
Exchanging the $\delta_i$ in favour of
\[\label{defnofdeltaij}
2\alpha_{ij}=\frac{\Delta_t}{3}-\Delta_i-\Delta_j,
\]
which is a solution of \eqref{deltaijcond_a}, 
the result is 
\begin{align}\label{I4res1}
\mathcal{I}_4 &= 2^4 (2\pi)^{-3d/2} \hat{C}_4\,\Big(\prod_{i=1}^3\int\D^d\bs{q}_i \Big) q_1^{\alpha_{14}-\alpha_{23}}
q_2^{-2\alpha_{24}-4\alpha_{13}-3d} q_3^{\alpha_{34}-\alpha_{12}}
\nn\\ &\qquad \times
 |\p_1+\q_2-\q_3|^{\alpha_{23}-\alpha_{14}}|\p_2+\q_3-\q_1|^{-2\alpha_{13}-4\alpha_{24}-3d}|\p_3+\q_1-\q_2|^{\alpha_{12}-\alpha_{34}}
\nn\\[1ex]&\qquad \times 
\int_0^\infty \D z \, z^{3d-\Delta_t/2-1}K_{\alpha_{23}-\alpha_{14}}(z\sqrt{\hat{u}})K_{\alpha_{13}-\alpha_{24}}(z)K_{\alpha_{12}-\alpha_{34}}(z/\sqrt{\hat{v}}),
\end{align}
where $\hat{u}$ and $\hat{v}$ are given in \eqref{uvhatdef}. 
As required, \eqref{I4res1} now has the form of a simplex integral \eqref{kint4}, namely
\begin{align}
& \mathcal{I}_4
=\int\frac{\D^d\bs{q}_1}{(2\pi)^d}\frac{\D^d\bs{q}_2}{(2\pi)^d}\frac{\D^d\bs{q}_3}{(2\pi)^d}\frac{\hat{f}(\hat{u},\hat{v})}{\mathrm{Den}_4(\bs{q}_j,\bs{p}_k)}.
\end{align}
The momentum routing through the tetrahedron is as illustrated in figure \ref{fig:1} and
the denominator $\mathrm{Den}_4(\bs{q}_j,\bs{p}_k)$ is given in \eqref{Den3}.  The function in the numerator is
\begin{align}
\hat{f}(\hat{u},\hat{v}) &= \tilde{C}_4
\Big(\frac{\hat{u}}{\hat{v}}\Big)^{(\alpha_{12}+\alpha_{34}+d)/2}
\nn\\ &\quad \times
 \int_0^\infty\D z \,
z^{3(\alpha_{12}+\alpha_{34}+d)-1}K_{\alpha_{23}-\alpha_{14}}(z\sqrt{\hat{u}})
K_{\alpha_{13}-\alpha_{24}}(z)
K_{\alpha_{12}-\alpha_{34}}\Big(\frac{z}{\sqrt{\hat{v}}}\Big)\label{tripleKfhatrep}
\end{align}
where
\[
\tilde{C}_4 =  2^4 (2\pi)^{3d/2}\hat{C}_4 = 2^{3d-\Delta_t/2+3}\pi^{3d/2}\Gamma\Big(\frac{\Delta_t-d}{2}\Big)\prod_{j=1}^4\frac{1}{\Gamma(\Delta_j-d/2)}.
\]
This is the result we sought: the specific function of momentum-space cross ratios appearing in the simplex representation for the $4$-point contact Witten diagram, as given in \cite{Bzowski:2019kwd}.  
Curiously this $\hat{f}(\hat{u},\hat{v})$ involves the same integral of three Bessel functions, the triple-$K$ integral, as appears in the $3$-point function $\mathcal{I}_3$, though the arguments and parameters are now different.
Specifically,
\begin{align}\label{fhatres}
&\hat{f}(\hat{u},\hat{v}) = \tilde{C}_4\, \frac{\hat{u}^{\alpha_{14}+d/2}}{\hat{v}^{\alpha_{34}+d/2}}
I_{3(\alpha_{12}+\alpha_{34}+d)-1,\,\{\alpha_{23}-\alpha_{14},\,\alpha_{13}-\alpha_{24},\,\alpha_{12}-\alpha_{34}\}}(\sqrt{\hat{u}},1,\frac{1}{\sqrt{\hat{v}}})
\end{align}
where the triple-$K$ integral \cite{Bzowski:2013sza} is  
\[\label{tripleKdef}
I_{\alpha,\,\{\beta_1,\beta_2,\beta_3\}}(p_1,p_2,p_3) = \int_0^\infty \D z \, z^{\alpha} \prod_{j=1}^3 p_j^{\beta_j} K_{\beta_j}(p_j z).
\]
For half-integer indices the Bessel functions reduce to elementary functions and the integral is trivial, while for integer values a general reduction scheme is available \cite{Bzowski:2015yxv,Bzowski:2020lip}.  An explicit evaluation is also known in terms of the double hypergeometric function Appell $F_4$ \cite{Bzowski:2013sza}.
For comparison, 
from \eqref{nptcontact} the 3-point function is 
\[\label{usualtripleK}
\mathcal{I}_3 = \Big(\prod_{j=1}^3 \frac{2^{1-\Delta_j+d/2}}{\Gamma(\Delta_j-d/2)}\Big) \,I_{d/2-1,\,\{\Delta_1-d/2,\Delta_2-d/2,\Delta_3-d/2\}}(p_1,p_2,p_3),
\]
which is equivalent to a 1-loop triangle integral  as we showed in \eqref{I3tri}.

In fact, as we show in appendix \ref{app_Dfn}, the form of $\hat{f}(\hat{u},\hat{v})$ in \eqref{fhatres} is very similar to that of the function of cross ratios $f(u,v)$ for the 4-point contact diagram in {\it position space}, also known as the holographic $D$-function \cite{DHoker:1999kzh}.  We will return to explain the origin of this connection, as well as its generalisation to the $n$-point contact diagram,  at the end of the following section.

\subsection{\texorpdfstring{Evaluating the $n$-point function via recursive convolutions}{Evaluating the n-point function via recursive convolutions}}\label{recursive_sec}

We now turn to 
the general $n$-point contact diagram for $n\ge 4$.  Once again, our goal is to find the specific function of momentum-space cross ratios appearing in the simplex representation \eqref{simplex} for this diagram.  
While the star-mesh approach above can likely be generalised in a recursive manner, 
we present instead an alternative method based on Fourier transforming the position space contact diagram using the convolution theorem.  These convolutions also have a recursive structure enabling their evaluation for general $n$.

In position space, the $n$-point contact diagram evaluates to 
\[
\mathcal{I}_n = \int\frac{\D z}{z^{d+1}}\int \D^d \bs{x}_0 \,\prod_{i=1}^n C_{\Delta_i}\Big(\frac{z}{z^2+x_{i0}^2}\Big)^{\Delta_i}, \label{posspcontact}
\]
where $\x_{ij}=\x_i-\x_j$ and the holographic normalisation is
\[
C_{\Delta_i} = \frac{\Gamma(\Delta_i)}{\pi^{d/2}\Gamma(\Delta_i-d/2)}.\label{CDeltadef}
\]
Schwinger parametrising all denominators and performing the $z$ integral, we find 
\begin{align}
\mathcal{I}_n &= \frac{1}{2}\Gamma\Big(\frac{\Delta_t-d}{2}\Big) \Big(\prod_{i=1}^n\frac{C_{\Delta_i}}{\Gamma(\Delta_i)}\int_0^\infty \D s_i  \,s_i^{\Delta_i-1} \Big) s_t^{(d-\Delta_t)/2} \int \D^d \bs{x}_0\, \exp\Big(-\sum_i s_i x_{i0}^2\Big).
\end{align}
To evaluate the $\bs{x}_0$ integral we complete
the square, expressing the remainder as
\begin{align}\label{compsq1}
\frac{1}{s_t}\Big(\sum_i s_i \bs{x}_i\Big)^2 - \sum_i s_i x_i^2 &= \frac{1}{s_t}\Big(\sum_i s_i(s_i-s_t)x_i^2 + 2\sum_{i<j} s_i s_j\, \bs{x}_i\cdot\bs{x}_j\Big) \nn\\ &
=\frac{1}{s_t}\Big(-\sum_{i<j} s_is_j(x_i^2+x_j^2) + 2\sum_{i<j} s_i s_j\, \bs{x}_i\cdot\bs{x}_j\Big) \nn\\ &
= -\frac{1}{s_t}\sum_{i<j}s_is_j x_{ij}^2,
\end{align}
after which we find
\begin{align}\label{step1}
\mathcal{I}_n &=  C_n\,\Big(\prod_{i=1}^n \int_0^\infty \D s_i\, s_i^{\Delta_i-1} \Big) s_t^{-\Delta_t/2}  \exp\Big(-\frac{1}{s_t}\sum_{i<j}s_is_j x_{ij}^2\Big)
\end{align}
where
\[
C_n = \frac{\pi^{d/2}}{2} \Gamma\Big(\frac{\Delta_t-d}{2}\Big) \prod_{i=1}^n \frac{C_{\Delta_i}}{\Gamma(\Delta_i)}
\]

As noted by  Symanzik in \cite{Symanzik:1972wj}, the value the integral \eqref{step1} is unchanged by making the replacement
\[
s_t =\sum_{i=1}^n s_i \rightarrow \sum_{i=1}^n \kappa_i s_i
\]
for any arbitrary set of $\kappa_i\ge 0$ not all zero.  (For a derivation, see appendix \ref{Symanzik_trick_appendix}.)
This observation is significant since by replacing $s_t\rightarrow s_1$ the integrand of \eqref{step1} becomes a product of exponential factors with a recursive structure:
\[\label{Schwingerstar}
\mathcal{I}_n =  C_n\,\Big(\prod_{i=1}^n \int_0^\infty \D s_i\, s_i^{\Delta_i-1} \Big) s_1^{-\Delta_t/2}  g_n
\]
where
\[
g_n = \prod_{1\le i<j}^n  \exp\Big(-\frac{s_is_j}{s_1}x_{ij}^2
\Big)=g_{n-1}\times \prod_{i=1}^{n-1} \exp\Big(-\frac{s_i s_n }{s_1}x_{in}^2\Big).
\]
In particular, the integrand $g_{n-1}$ has no dependence on either $\x_n$ or $s_n$ (and it is this latter property which required the replacement $s_t\rightarrow s_1$). 
Upon Fourier transforming, this recursive product
becomes a convolution,
\begin{align}
&\mathcal{F}[g_n](\p_1,\ldots,\p_n) =\Big( \mathcal{F}[g_{n-1}](2\pi)^d\delta(\p_n)\Big)\ast \mathcal{F}\Big[\prod_{i=1}^{n-1} \exp\Big(-\frac{s_i s_n }{s_1}x_{in}^2\Big)\Big],
\end{align}
where the $\delta(\p_n)$ arises since $g_{n-1}$ is independent of $\x_n$. 
Explicitly, we find
\begin{align}
\mathcal{F}[g_n](\p_1,\ldots,\p_n)&
=
\Big(\prod_{k=1}^n\int\frac{\D^d \q_{k}}{(2\pi)^d}  \Big)\,\mathcal{F}[g_{n-1}](\p_1-\q_1,\ldots, \p_{n-1}-\q_{n-1})(2\pi)^d\delta(\p_n-\q_n)\nn\\&\qquad \qquad\times (2\pi)^d\delta\Big(\sum_{j=1}^n \q_j\Big) \prod_{i=1}^{n-1} \Big(\frac{\pi s_1}{s_n s_i}\Big)^{d/2}\exp\Big(-\frac{s_1 q_{i}^2}{4s_i s_n }\Big)\nn\\[1ex]&=
\prod_{i=1}^{n-1}\Big(\int\frac{\D^d \q_{in}}{(2\pi)^d} \,\Big(\frac{\pi s_1}{s_n s_i}\Big)^{d/2}\!\exp\Big(-\frac{s_1 q_{in}^2}{4s_i s_n }\Big)\Big) (2\pi)^d\delta\Big(\p_n+\sum_{j=1}^{n-1} \q_{jn}\Big)  \nn\\&\qquad \qquad\times  \mathcal{F}[g_{n-1}](\p_1-\q_{1n},\ldots, \p_{n-1}-\q_{n-1,n}),
\end{align}
where  in the second equation we recognise the integration momenta as $\q_{in}$, the momenta running from vertex $i$ to vertex $n$.
To complete our evaluation, we can now exploit the recursive structure.
Starting with $g_1=1$, we find
\begin{align}
\mathcal{F}[g_1](\p_1) &= (2\pi)^d\delta(\p_1),\\[1ex]
\mathcal{F}[g_2](\p_1,\p_2) &= \int\frac{\D^d \q_{12}}{(2\pi)^d}\,
\Big(\frac{\pi}{s_2}\Big)^{d/2}\!\exp\Big(-\frac{q_{12}^2}{4s_2}\Big)
(2\pi)^d\delta(\p_1+\q_{21})(2\pi)^d\delta(\p_2+\q_{12})
\end{align}
and hence, after $n$ iterations, 
\begin{align}
&\mathcal{F}[g_n](\p_1,\ldots,\p_n)\nn \\&\quad = \Big(\prod_{1\le i<j}^n\int\frac{\D^d\q_{ij}}{(2\pi)^d}\,\Big(\frac{\pi s_1}{s_i s_j}\Big)^{d/2}\! \exp\Big(-\frac{s_1q_{ij}^2}{4s_is_j}\Big)\Big)
\prod_{k=1}^n (2\pi)^d\delta\Big(\p_k+\sum_{l=1}^n \q_{lk}\Big).
\end{align}
Restoring the Schwinger integrations from \eqref{Schwingerstar}, we see the momentum-space contact diagram $\mathcal{I}_n$ thus has the expected structure of a simplex integral \eqref{simplex}
with 
\begin{align}
\hat{f}_n(\hat{\bs{u}}) = C_n\Big(\prod_{k=1}^n \int_0^\infty \D s_k\, s_k^{\Delta_k-1}\Big) s_1^{-\Delta_t/2}  \prod_{1\le i<j}^n\Big(\frac{\pi s_1}{s_i s_j}\Big)^{d/2} \exp\Big(-\frac{s_1q_{ij}^2}{4s_is_j}\Big)q_{ij}^{2\alpha_{ij}+d}
\end{align}
where $\alpha_{ij}$ satisfies \eqref{deltaijcond_a}.   As we will show shortly, this result is indeed a function of only the momentum-space cross ratios $\hat{\bs{u}}$. 
First, however, we 
note that a fully symmetric parametrisation can be recovered via another application of the Symanzik trick.  
To see this, we substitute $s_i = 1/t_i$ yielding
\begin{align}\label{tform1}
\hat{f}_n(\hat{\bs{u}}) = C_n\Big(\prod_{k=1}^n \int_0^\infty \D t_k\, t_k^{-\Delta_k-1}\Big) t_1^{\Delta_t/2}  \prod_{1\le i<j}^n\Big(\frac{\pi t_i t_j}{t_1}\Big)^{d/2} \exp\Big(-\frac{t_i t_j q_{ij}^2}{4t_1}\Big)q_{ij}^{2\alpha_{ij}+d}
\end{align}
Substituting $t_i=\sigma y_i$ 
and following 
the steps described in appendix \ref{Symanzik_trick_appendix}, we find
\begin{align}\label{alphaform1}
\hat{f}_n(\hat{\bs{u}}) &= C_n \,2^{-\Delta_t} (4\pi)^{n(n-1)d/4}\Gamma\Big(n(n-1)\frac{d}{4}-\frac{\Delta_t}{2}\Big)\Big(\prod_{1\le i<j}^n q_{ij}^{2\alpha_{ij}+d}\Big)
\nn\\&\quad \times \Big(\prod_{k=1}^n \int_0^1 \D y_k \, y_k^{(n-1)d/2-\Delta_k-1}\Big) \delta\big(1-\sum_{i=1}^n\kappa_i y_i\Big)\Big(\sum_{1\le i<j}^n y_i y_j q_{ij}^2\Big)^{\Delta_t/2-n(n-1)d/4}.
\end{align}
However, as also discussed in appendix \ref{Symanzik_trick_appendix}, 
this expression is equal to 
\begin{align}\label{tform2}
\hat{f}_n(\hat{\bs{u}}) = C_n\Big(\prod_{k=1}^n \int_0^\infty \D t_k\, t_k^{-\Delta_k-1}\Big) t_T^{\Delta_t/2}  \prod_{1\le i<j}^n\Big(\frac{\pi t_i t_j}{t_T}\Big)^{d/2} \exp\Big(-\frac{t_i t_j q_{ij}^2}{4t_T}\Big)q_{ij}^{2\alpha_{ij}+d}
\end{align}
where $t_T = \sum_{i=1}^n \kappa_i t_i$ for any $\kappa_i\ge 0$ not all zero.  Choosing $\kappa_{i} =\delta_{i1}$ then corresponds to \eqref{tform1}, while setting all the $\kappa_i$ to unity yields a fully symmetric parametrisation.
This fully symmetric parametrisation has a very similar form  to 
our original position-space integral \eqref{step1} (modulo the factors of $q_{ij}^{2\alpha_{ij}+d}$, but corresponding factors of the coordinate separations can be introduced in \eqref{step1} by considering instead the function $f_n(\bs{u})$.)
However, the internal momenta $\q_{ij}$ are not in general differences as was the case for the  $\x_{ij}=\x_i-\x_j$.

\begin{figure}[t]
\begin{center}
\includegraphics[width=5.5cm]{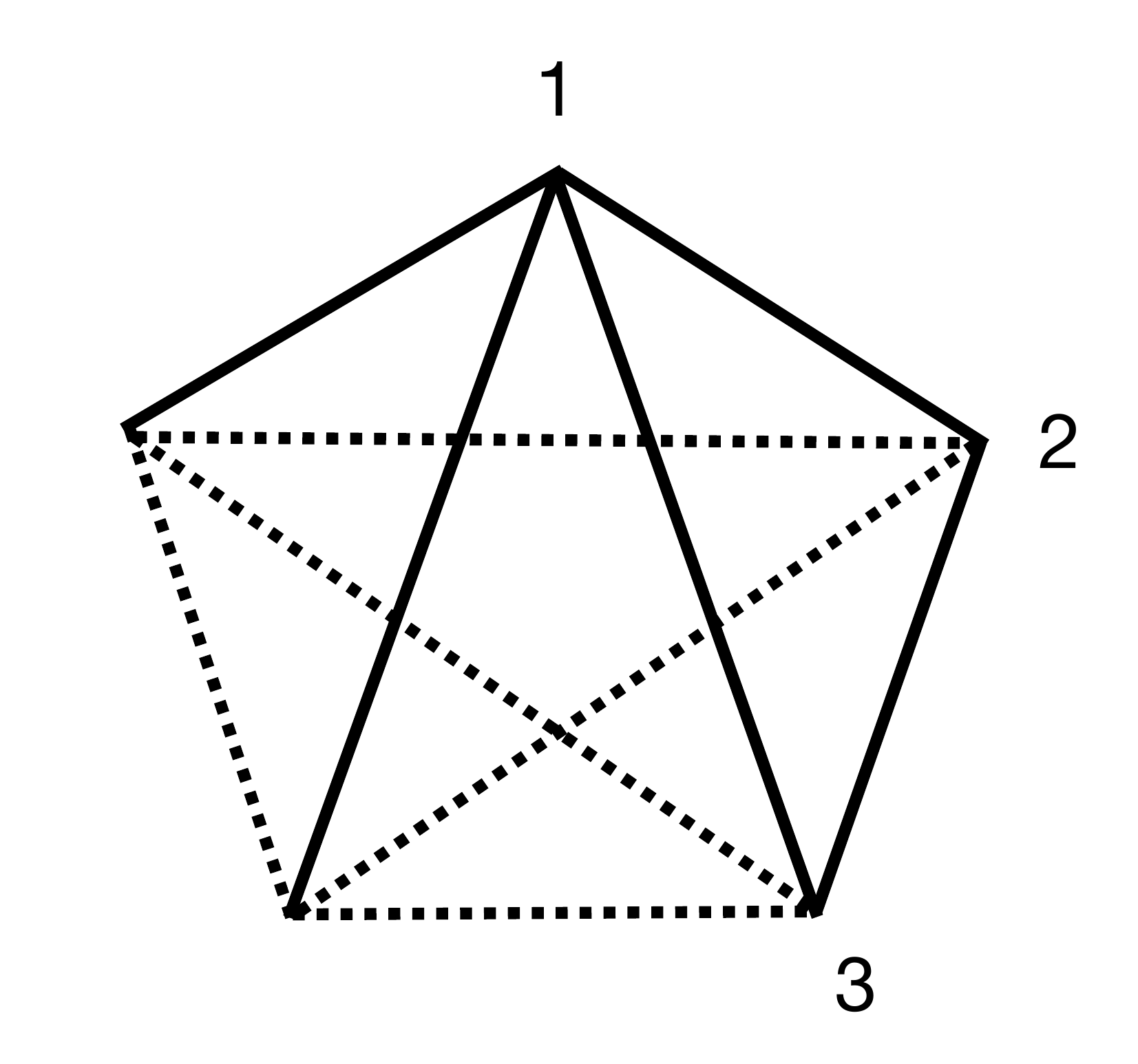}
\end{center}
\caption{In \eqref{newparam}, we exchange the Schwinger parameters $t_i$ where $i=1,\ldots,n$ for a new set consisting of $z_{23}$ and $z_{1i}$ for $i=2,\ldots n$.  These correspond to the solid legs on the diagram above, shown for the case $n=5$.
\label{legfig}}
\end{figure}

Let us now show that this $\hat{f}_n(\hat{\bs{u}})$ is indeed a function of only the momentum-space cross ratios.  A convenient way to see this is to 
select new independent variables corresponding to the subset of $n$ legs shown in figure \ref{legfig}.  We thus  
parametrise 
\[\label{newparam}
t_1 =\frac{z_{12}z_{13}}{z_{23}}\frac{q_{23}^2}{q_{12}^2 q_{13}^2}, \qquad t_i = \frac{z_{1i}}{q_{1i}^2}, \qquad i=2,\ldots, n
\]
and introduce the $n(n-3)/2$ independent momentum-space cross ratios 
\begin{align}
\hat{u}_{2a} = \frac{q_{2a}^2 q_{13}^2}{q_{1a}^2 q_{23}^2},  \qquad
\hat{u}_{3a} = \frac{q_{3a}^2 q_{12}^2}{q_{1a}^2 q_{23}^2}, \qquad \hat{u}_{ab} = \frac{q_{ab}^2 q_{23}^2}{q_{2a}^2 q_{3b}^2}\label{fullcrossratios}
\end{align}
where $a,b=4,\ldots,n$ and in the last equation $a<b$ with no sum implied.
Converting \eqref{tform1} into these new variables, we find
\begin{align}\label{tform3}
\hat{f}_n(\hat{\bs{u}})& = C_n \pi^{n(n-1)d/4}\Big(\prod_{1\le k<l}^n q_{kl}^{2\alpha_{kl}+d}\Big)
\Big(\prod_{i=2}^n \int_0^\infty 
\frac{\D z_{1i}}{z_{1i}}
\Big(\frac{z_{1i}}{q_{1i}^2}\Big)^{-\Delta_i+(n-1)d/2} \Big)\nn\\&\quad \times 
\int_0^\infty \frac{\D z_{23}}{z_{23}}\Big( \frac{z_{12}z_{13}}{z_{23}}\frac{q_{23}^2}{q_{12}^2 q_{13}^2}\Big)^{\Delta_t/2-\Delta_1-(n-1)(n-2)d/4}
\nn\\& \quad \times
\exp\Big[-\frac{1}{4}\Big(z_{12}+z_{23}+z_{13}+\sum_{a=4}^n z_{1a}\big(1+\frac{z_{23}}{z_{13}}\hat{u}_{2a}+\frac{z_{23}}{z_{12}}\hat{u}_{3a}\big)\nn\\&\qquad \qquad
+\sum_{4\le a<b}^n \frac{z_{1a}z_{1b}z_{23}}{z_{12}z_{13}}\hat{u}_{ab}\hat{u}_{2a}\hat{u}_{3b}\Big)\Big].
\end{align}
To cancel the remaining $q_{ij}^2$, let us then choose 
\begin{align}\label{deltasols1}
\alpha_{12} &= -\Delta_1 -\Delta_2 + \frac{\Delta_t}{2}-(n-2)(n-3)\frac{d}{4}, \\
\alpha_{13} &= -\Delta_1 -\Delta_3 + \frac{\Delta_t}{2}-(n-2)(n-3)\frac{d}{4}, \\
\alpha_{23}&= \Delta_1 -\frac{\Delta_t}{2}+n(n-3)\frac{d}{4},\\
\alpha_{1a}&= -\Delta_a+(n-2)\frac{d}{2},\\
\alpha_{2a}&=\alpha_{3a}=-\frac{d}{2}, \\
\alpha_{ab}&=-\frac{d}{2}\label{deltasols2}
\end{align}
where again $a,b = 4,\ldots,n$ and $a<b$.  One can check that this choice is a valid solution of the constraint \eqref{deltaijcond_a}.\footnote{Making a different choice here multiplies $\hat{f}_n(\hat{\bs{u}})$ by powers of the cross ratios, but this change is off-set by the different powers that will then appear in the denominator of the simplex representation.  Indeed, for $n=4$,  \eqref{deltasols1}-\eqref{deltasols2} represent a different choice to \eqref{defnofdeltaij} earlier, which we used for compatibility with \cite{Bzowski:2019kwd}.} 
We thus arrive at the following expression for $\hat{f}_n(\hat{\bs{u}})$ which is manifestly a function of the momentum-space cross ratios only,
\begin{align}\label{finalform1}
\hat{f}_n(\hat{\bs{u}})& = C_n \pi^{n(n-1)d/4}
\Big(\prod_{i=2}^n \int_0^\infty 
\D z_{1i}\,z_{1i}^{\alpha_{1i}+d/2-1} \Big)
\int_0^\infty \D z_{23}\, z_{23}^{\alpha_{23}+d/2-1}
\nn\\& \quad \times
\exp\Big[-\frac{1}{4}\Big(z_{12}+z_{23}+z_{13}+\sum_{a=4}^n z_{1a}\big(1+\frac{z_{23}}{z_{13}}\hat{u}_{2a}+\frac{z_{23}}{z_{12}}\hat{u}_{3a}\big)\nn\\&\qquad \qquad
+\sum_{4\le a<b}^n \frac{z_{1a}z_{1b}z_{23}}{z_{12}z_{13}}\hat{u}_{ab}\hat{u}_{2a}\hat{u}_{3b}\Big)\Big].
\end{align}

We can simplify this expression 
still further by performing another two integrations.
Evaluating the $z_{23}$ integral, we find
\begin{align}
\hat{f}_n(\hat{\bs{u}})& = C_n 2^{2\alpha_{23}+d}\pi^{n(n-1)d/4}\Gamma\big(\alpha_{23}+\frac{d}{2}\big)
\Big(\prod_{i=2}^n \int_0^\infty 
\D z_{1i}\,z_{1i}^{\alpha_{1i}+d/2-1} e^{-z_{1i}/4}\Big) (z_{12}z_{13})^{\alpha_{23}+d/2}
\nn\\& \quad \times
\Big(z_{12}z_{13}+
\sum_{a=4}^n z_{1a}\big(z_{12}\hat{u}_{2a}+z_{13}\hat{u}_{3a}\big)
+\sum_{4\le a<b}^n z_{1a}z_{1b}\hat{u}_{ab}\hat{u}_{2a}\hat{u}_{3b}\Big)^{-\alpha_{23}-d/2}.
\end{align}
We now set
$
z_{1i} = \sigma y_{1i}
$
for $i=2,\ldots, n$, subject to the constraint
$\sum_{i=2}^n y_{1i}= 1$ so that the exponential reduces to $e^{-\sigma/4}$.  The Jacobian can be evaluated as discussed in appendix \ref{Symanzik_trick_appendix} and the $\sigma$ integral performed, taking us to
\begin{align}\label{optform}
\hat{f}_n(\hat{\bs{u}})& =\hat{C}_n 
\Big(\prod_{i=2}^n \int_0^1
\D y_{1i}\,y_{1i}^{\alpha_{1i}+d/2-1} \Big) (y_{12}y_{13})^{\alpha_{23}+d/2}\delta\Big(1-\sum_{i=2}^n y_{1i}\Big)
\nn\\& \quad \times
\Big(y_{12}y_{13}+
\sum_{a=4}^n y_{1a}\big(y_{12}\hat{u}_{2a}+y_{13}\hat{u}_{3a}\big)
+\sum_{4\le a<b}^n y_{1a}y_{1b}\hat{u}_{ab}\hat{u}_{2a}\hat{u}_{3b}\Big)^{-\alpha_{23}-d/2}
\end{align}
where  the $\alpha_{ij}$ are given by \eqref{deltasols1}-\eqref{deltasols2}, and the normalisation is 
\[
\hat{C}_n =  C_n \pi^{n(n-1)d/4}4^{nd/2+\alpha_{23}-\Delta_1} \Gamma\Big(\alpha_{23}+\frac{d}{2}\Big)\Gamma\Big( (n-1)\frac{d}{2}-\Delta_1\Big)
\]
where we used \eqref{deltaijcond_a} to replace $\sum_{i=2}^n\alpha_{1i}=-\Delta_1$.
We retain the overall delta function in \eqref{optform} for symmetry, but after its removal only $(n-2)$ integrations remain.
This appears to be the optimal representation for $\hat{f}_n(\hat{\bs{u}})$.  For comparison, the Mellin-Barnes representation obtained following Symanzik's procedure in \cite{Symanzik:1972wj} has $n(n-3)/2$ Mellin integrations which is larger than $n-2$ for any $n> 4$.

For the $4$-point function, \eqref{optform} reduces to 
\begin{align}\label{4optform}
\hat{f}_4(\hat{\bs{u}})& =\hat{C}_4
 \int_0^1
\D y_{12} \,y_{12}^{\alpha_{12}+\alpha_{23}+d-1}\int_0^1
\D y_{13}\, y_{13}^{\alpha_{13}+\alpha_{23}+d-1}\int_0^1 
\D y_{14}\,y_{14}^{\alpha_{14}+d/2-1} 
\nn\\& \quad \times\delta\big(1-y_{12}-y_{13}-y_{14}\big)
\Big(y_{12}y_{13}+
 y_{12}y_{14}\hat{u}_{24}+y_{13}y_{14}\hat{u}_{34}\Big)^{-\alpha_{23}-d/2}.
\end{align}
This is precisely the Feynman parametrisation of the 1-loop triangle integral (see, {\it e.g.,} appendix A.3 of \cite{Bzowski:2013sza}), except that the momentum-space cross ratios $\hat{u}_{24}$ and $\hat{u}_{34}$ take the place of ratios of the squared external momenta.  This makes sense since in section \ref{startrisection} we showed that the 1-loop triangle is equivalent to a triple-$K$ integral, and from \eqref{tripleKfhatrep} we know that $\hat{f}_4(\hat{\bs{u}})$ can be written as a triple-$K$ integral where the arguments are given by the momentum-space cross ratios.
The exact equivalence of \eqref{4optform} and \eqref{tripleKfhatrep} can be verified using equation (A.3.23) of \cite{Bzowski:2013sza}, taking into account the additional powers of $\hat{u}$ and $\hat{v}$ that arise from reparameterising  the denominator of the simplex in terms of the $\alpha_{ij}$ in \eqref{defnofdeltaij} instead of those in \eqref{deltasols1}-\eqref{deltasols2}.

In summary then, the $n$-point contact diagram can indeed be written as a simplex integral.  Equation \eqref{optform} expresses $\hat{f}_n(\hat{\bs{u}})$, the corresponding function of momentum-space cross ratios,  as an $(n-2)$-fold  Feynman parametric integral over a quadratic denominator.
To finish, let us point out a few of the remarkable connections between this result and other Feynman integrals.

Firstly, there is a close similarity between equation \eqref{optform} for $\hat{f}_n(\hat{\bs{u}})$ and the corresponding representation for $f_n(\bs{u})$, the function of cross ratios parametrising the $n$-point contact diagram in position space: repeating the steps from \eqref{newparam} to \eqref{optform} above, but starting instead from \eqref{step1}, we find the position-space contact diagram is
\[\label{Inpossp}
\mathcal{I}_n = \prod_{1\le i<j<n} x_{ij}^{-2\alpha_{ij}-d} f_n(\bs{u})
\]
where $f_n(\bs{u})$ is now given by  {\it precisely} the right-hand side of \eqref{optform}.   The $\alpha_{ij}$ are given by \eqref{deltasols1}-\eqref{deltasols2} after making the replacements $\Delta_i\rightarrow -\Delta_i$ and setting $n\rightarrow 1$ in these formulae.  These new $\alpha_{ij}$ do not satisfy \eqref{deltaijcond_a}, but this is correct since from \eqref{Inpossp} the necessary condition is instead $\sum_{j\neq i}(\alpha_{ij}+d/2)=\Delta_i$, which is satisfied.
In addition, we replace $\hat{C}_n \rightarrow C_n \Gamma(\Delta_1)\Gamma(\alpha_{23}+d/2)$ and $\hat{\bs{u}}\rightarrow \bs{u}$ by sending $q_{ij}^2\rightarrow x_{ij}^2$ in \eqref{fullcrossratios}.   
Thus, modulo these replacements, both $\hat{f}_n(\hat{\bs{u}})$ and $f_n(\bs{u})$ have exactly the same parametrisation \eqref{optform}.
In other words, the function $\hat{f}_n(\hat{\bs{u}})$ of  {\it momentum-space} cross ratios appearing in the simplex representation has the same form as  the function $f_n(\bs{u})$ of {\it ordinary} cross ratios describing the contact diagram in position space. 

The source of this surprising equivalence is easier to see by comparing \eqref{step1} (or rather, the corresponding $f_n(\bs{u})$) with \eqref{tform2}.
The Fourier transform of a product of Gaussians is a convolution of Gaussians.  Rewriting this convolution as a simplex, the  resulting $\hat{f}_n(\hat{\bs{u}})$ in  \eqref{tform2} is identical to our starting point \eqref{step1} up a change of parameters.

This connection also explains the curious resemblance of \eqref{alphaform1} and \eqref{optform} to massless $n$- and $(n-1)$-point polygon Feynman integrals respectively.
The $n$-point contact diagram in position space can be rewritten as a conformal Symanzik star integral living in spacetime dimension $\Delta_t$ \cite{Dolan:2000ut}.  In the dual momentum  (or region) coordinates, this becomes a conformal $n$-point polygon integral, whose Feynman-parametrised representation is equivalent to \eqref{alphaform1}.   Performing further integrations as above, we arrive at the representation \eqref{optform} whose structure corresponds to an $(n-1)$-point  polygon but with Lorentz invariants of the external momenta replaced by momentum-space cross ratios.  
For the 4-point function, this corresponds to the observation discussed in appendix \ref{app_Dfn}: the contact diagram (or holographic `$D$-function') 
  is equivalent to a conformal box integral in spacetime dimension $\Delta_t$, which is  in turn  equivalent to a 1-loop triangle or triple-$K$ integral where the external momenta are replaced by cross ratios.

\section{Discussion}
\label{Discussionsec}

In this work, we proved that the general $n$-point function of any CFT can be expressed as a generalised Feynman integral over an $(n-1)$-simplex in momentum space.  This integral features an arbitrary function of the $n(n-3)/2$ independent cross ratios constructed from the momenta running between the vertices of the simplex.   The invariance of general simplex integrals is closely related to the invariance of the more specialised mesh integrals, whose recursive structure enables a simple generalisation of results to arbitrary $n$.

Given any conformal correlator has a simplex representation, an immediate task is then to construct this representation explicitly for known $n$-point correlators.
Perturbative examples were studied in \cite{Bzowski:2019kwd}, and here we derived the simplex representation for the $n$-point contact diagram of a general holographic CFT. 
We showed the relevant function of momentum-space cross ratios can be written as an $(n-2)$-fold Feynman parametrised integral, whose form is identical (up to certain changes of parameters) to that of the function of cross ratios associated with the original position-space contact diagram.  For the 4-point function, this function can conveniently be expressed in terms of a triple-$K$ integral \cite{Bzowski:2019kwd}.

The methods we developed for this analysis -- based on the star-mesh duality of electrical circuit theory and a recursive application of the convolution theorem -- should both be applicable to a wider class of examples.  In particular, the latter method requires only that correlators have a recursive product structure in position space, which can generally be arranged through a suitable Schwinger or Feynman parametrisation.
More broadly, a general recursive approach to the simplex might provide an alternative to use of the OPE or conformal block decomposition.  
A number of recursive methods have been proposed for holographic CFTs (see, {\it e.g.,} \cite{Raju:2010by, Raju:2012zr,Arkani-Hamed:2017fdk, Albayrak:2019asr,Zhou:2020ptb}), 
but one might ask quite generally whether higher-point simplices 
can be decomposed in terms of lower-point ones.

Another attractive  target for further investigation are holographic exchange diagrams.  Many specific momentum-space examples, including exchanges of fields with nonzero spin, have recently been constructed by acting on known seed correlators with spin- and weight-raising operators \cite{Arkani-Hamed:2015bza, Arkani-Hamed:2018kmz, Baumann:2019oyu, Baumann:2020dch, Sleight:2019hfp, Sleight:2019mgd, Sleight:2020obc}.  Still more have been constructed following the methods of \cite{Raju:2012zs}, including examples at 5- and 6-points \cite{Albayrak:2018tam,Albayrak:2019yve, Albayrak:2020isk}.
It would be interesting to determine the simplex representation for some of  these, particularly those with cosmological relevance.  Given their low transcendality,  we expect such examples to be associated with reductions in the loop order of the simplex integral.  This can occur through at least two distinct  mechanisms as discussed in \cite{Bzowski:2019kwd}.  
Beyond these examples, it would  be desirable to understand the simplex representation for  general exchange diagrams involving operators of arbitrary dimension and spin.  
And beyond holographic CFTs, we wish to construct a simplex representation for general $n$-point correlators involving tensorial external operators, valid in any CFT.  This may be achievable through the action of spin-raising operators, or else through the development of a form factor decomposition extending that for 3-point functions in \cite{Bzowski:2013sza}. For work at 4-points, see  \cite{Dymarsky:2013wla, Coriano:2019nkw,Maglio:2019grh, Serino:2020pyu}.

Other natural questions include the following.
Can the simplex representation be used to understand the singularities of conformal correlators?  An classification of  divergences for the 4-point function was made in \cite{Bzowski:2019kwd}, but the detailed structure of conformal anomalies and the renormalisation of higher-point correlators remains to be explored.
On a similar note, 
 how do the factorisation properties of momentum-space correlators envisaged by Polyakov \cite{Polyakov:1974gs} arise?  Related recent discussions include  \cite{Sen:2015doa, Gopakumar:2016wkt, Gopakumar:2016cpb, Gopakumar:2018xqi, Isono:2018rrb, Isono:2019wex, Sleight:2019ive}.  Since the simplex representation is a generalised Feynman integral, it should be especially well-suited for extracting the discontinuities needed to understand the implications of  unitarity.
 
Next, how can we analytically continue our results from Euclidean to Lorentzian signature?  For 3-point functions this has been analysed recently in  \cite{Bautista:2019qxj,Gillioz:2019lgs,Anand:2019lkt}, but what happens for the simplex?  
Furthermore, the flat-space limit of CFT correlators yields  scattering amplitudes \cite{Polchinski:1999ry, Gary:2009ae, Penedones:2010ue}; for momentum-space discussions see 
 \cite{Raju:2012zr, Farrow:2018yni, Lipstein:2019mpu}.  
 How are these scattering amplitudes then related to the arbitrary function of momentum-space cross ratios appearing in the simplex representation?
Such an understanding would clarify the 
structure of correlators, both for cosmology \cite{Arkani-Hamed:2018kmz} and in relation to the S-matrix bootstrap \cite{Gillioz:2020mdd}.

There are also  questions of a more mathematical nature.  Are there any  applicable results from graph theory or simplex homology?
In topology, one may define a boundary map $\partial_n$ that acts on oriented $n$-simplices yielding a linear combination of $(n-1)$-simplices, and squares to zero, {\it i.e.}  $\partial_{n-1} \partial_n=0$.  We have already seen that there is a natural recursive structure associated with CFT correlators. Could this  be related to the boundary map, or more generally, can the boundary map be  extended to act on the CFT $n$-point function? Similarly, one may define graph Laplacians  acting on simplices: are these natural objects acting on the CFT? 

A connection to Grassmannians also seems likely, given that CFT correlators are mapped by the flat-space limit  to scattering amplitudes \cite{Arkani-Hamed:2018ign, GrassmanniansBook}.
In fact,  the relation  \eqref{Kleinquadric} obeyed by the currents on the tetrahedron also has links 
to the Klein correspondence, an important progenitor of twistor theory.
In the original context of electrical circuits where the currents are treated as scalars, this relation 
defines the Klein quadric in projective 5-space, with  the six currents representing homogeneous Pl{\"u}cker coordinates.  
According to the Klein correspondence \cite{KleinCorr1, KleinCorr2}, points on this quadric correspond to lines in projective 3-space.
The mapping is given by the Pl{\"u}cker embedding, which corresponds to  the star-mesh transform \eqref{starmesh1}.
Taken together, the set of all possible lines in projective 3-space is a manifold, the Grassmannian $Gr(2,4)$.  
Can this connection  be generalised to the full vectorial setting implicit in \eqref{starmesh1} and to arbitrary $n$, and if so, what are its physical implications?   

Finally, an exciting application of the simplex representation is to formulate  momentum-space approaches to the conformal bootstrap.  As a first step, one might try to understand the expansion of the $n$-point functions in conformal partial waves \cite{Ferrara:1973vz, Ferrara:1973yt, Polyakov:1974gs, Dolan:2003hv, Dolan:2011dv} and to identify the simplex representation for conformal blocks. Applying the momentum-space conformal Casimir operator to the simplex, we can extract the resulting partial differential equation obeyed by the arbitrary function of the momentum-space cross ratios.  Solving this equation would yield the eigenfunctions of the Casimir operator, and hence the conformal blocks \cite{Dolan:2003hv}; see also the recent works \cite{Gillioz:2018mto, Gillioz:2019iye, Gillioz:2020mdd}.  In addition, it would be useful to develop a better understanding of the momentum-space OPE limit in relation to the simplex, including a careful treatment of the short-distance singularities \cite{Bzowski:2014qja}.

Looking ahead, the simplex has many interesting facets 
worthy of further exploration.
We hope to shed light on some of these soon.

\section*{Acknowledgements}
AB is supported by the Knut and Alice Wallenberg Foundation under grant 113410212.
PLM is supported by the Science and Technology Facilities Council through an Ernest Rutherford Fellowship (ST/P004326/2).  
 KS is supported in part by the Science and Technology Facilities Council (Consolidated Grant “Exploring the Limits of the Standard Model and Beyond”).

\appendix

\section{Proofs of conformal invariance}
\label{app:confproofs}
\subsection{Recursive analysis for mesh integrals} \label{app:conf_mesh_P}

In the following we will  set up a recursive proof of the special conformal Ward identity \eqref{special} for mesh integrals \eqref{mesh}. We want to prove the recursive formula \eqref{recursiveE}
\begin{align}
\mathcal{E}^{(n) \kappa}_{SCWI}(\bs{\Delta}^{(n)}; \bs{p}_1, \ldots, \bs{p}_n )
&=\prod_{i = 1}^{n-1} C_{in} \int \frac{\D^d \bs{q}_{i}}{(2 \pi)^d} \frac{1}{q_1^{2 \alpha_{1n} + d} q_2^{2 \alpha_{2n} + d} \ldots q_{n-1}^{2 \alpha_{n-1,n} + d}}  \nn\\
& \quad\times (2 \pi)^d\delta\Big( \bs{p}_n + \sum_{j=1}^{n-1}\bs{q}_j\Big)
\mathcal{E}^{(n-1) \kappa}_{SCWI}(\bs{\Delta}^{(n-1)}; \bs{p}_1, \ldots, \bs{p}_{n-1}),
\end{align}
where $\mathcal{E}^{(n) \kappa}_{SCWI}$ is the action of the SCWI operator \eqref{special} on the mesh integral $M_n$ as defined in \eqref{meshSCWI}. For this purpose, we need a number of preliminary results.  
First, using the Fourier representation of the Dirac delta function and integration by parts, we find that
\[\label{trick2}
\K^\kappa(0;\bs{p}) (2\pi)^d \delta(\bs{p}) = 0,
\]
where the operator $\K^\kappa(\Delta; \bs{p})$ is defined in \eqref{Kopdef}.

Next, by direct differentiation, we obtain
\[\label{trick3}
\K^\kappa(\Delta;\bs{p}) (p^{2\Delta-d} F) = p^{2\Delta-d} \K^\kappa(d-\Delta;\bs{p})F, 
\]
where  the dimension inside the operator on the right-hand side is replaced by its shadow.
Using this relation, we can now establish the following integration by parts identity,
\begin{align}\label{trick1}
&\int \frac{\D^d\bs{q}}{(2\pi)^d}\frac{1}{q^{2\alpha+d}}F(\bs{q})\K^\kappa(\Delta;\bs{p})M(\bs{p}-\bs{q})
\nn\\& \quad = 
\int \frac{\D^d\bs{q}}{(2\pi)^d}\frac{1}{q^{2\alpha+d}}\Big[ F(\bs{q}) \K^\kappa(\Delta+\alpha;\bs{p}-\bs{q})M(\bs{p}-\bs{q})+M(\bs{p}-\bs{q})\K^\kappa(\alpha+d;\bs{q})F(\bs{q})\Big],
\end{align}
where $M$ and $F$ are arbitrary functions.
To verify this, we  check by direct calculation that
\[
\K^\kappa(\Delta; \bs{p})M(\bs{p}-\bs{q})=\Big(\K^\kappa(\Delta+\alpha;\bs{p}-\bs{q}) + \K^\kappa(\alpha+d;\bs{q})\Big)M(\bs{p}-\bs{q}) .
\]
The first term on this right-hand side gives the first term on the right-hand side of \eqref{trick1}.
The second term on this right-hand side can then be integrated by parts:
\begin{align}
&\int \frac{\D^d\bs{q}}{(2\pi)^d}\frac{1}{q^{2\alpha+d}}F(\bs{q})\K^\kappa(\alpha+d; \bs{q})M(\bs{p}-\bs{q})\nn\\
&=\int \frac{\D^d\bs{q}}{(2\pi)^d}M(\bs{p}-\bs{q})\Big(\frac{\partial}{\partial q^\nu} \frac{\partial}{\partial q^\nu}  q^\kappa -2\frac{\partial}{\partial q^\nu} \frac{\partial}{\partial q^\kappa} q^\nu-2\alpha \frac{\partial}{\partial q^\kappa} 
\Big)\frac{1}{q^{2\alpha+d}}F(\bs{q})\nn\\&
=\int \frac{\D^d\bs{q}}{(2\pi)^d}M(\bs{p}-\bs{q})\K^\kappa(-\alpha; \bs{q})\frac{1}{q^{2\alpha+d}}F(\bs{q})
\nn\\&
=\int \frac{\D^d\bs{q}}{(2\pi)^d}\frac{1}{q^{2\alpha+d}}M(\bs{p}-\bs{q})\K^\kappa(d+\alpha; \bs{q})F(\bs{q}),
\end{align}
where in the last line we used \eqref{trick3}.  This is now the second term on the right-hand side of \eqref{trick1} establishing this latter identity.

We now turn to the main proof.  For compactness, let us write
\begin{align}
\bs{p}_i' &= \bs{p}_i - \bs{q}_i, \qquad i=1,\ldots, n-1,\\
\bs{p}_n' &= \bs{p}_n + \sum_{j=1}^{n-1}\bs{q}_j,
\end{align}
and suppress the (fixed) $\alpha_{ij}$ indices inside $M_n$.
We then evaluate the action of the special conformal Ward identity on the recursive representation \eqref{meshrec0} of the mesh integral $M_n$.  Noting that $\p_n$ appears only inside the delta function $\delta(\p'_n)$, while the remaining external momenta appear only inside $M_{n-1}$, we have 
\begin{align}
\mathcal{E}^{(n)\kappa}_{SCWI}
& =\prod_{i = 1}^{n-1} C_{in} \int \frac{\D^d \bs{q}_{i}}{(2 \pi)^d} \frac{1}{q_1^{2 \alpha_{1n} + d} q_2^{2 \alpha_{2n} + d} \ldots q_{n-1}^{2 \alpha_{n-1,n} + d}}\nn\\&\qquad \times\Big[M_{n-1}(\bs{p}'_1, \ldots, \bs{p}'_{n-1}) \K^\kappa(\Delta_n^{(n)};\bs{p}_n)(2 \pi)^d \delta ( \bs{p}'_n) \nn\\&
\qquad\quad  +(2 \pi)^d \delta( \bs{p}'_n)
\Big(\sum_{m=1}^{n-1} \K^\kappa(\Delta_m^{(n)};\bs{p}_m) \Big)M_{n-1}(\bs{p}'_1, \ldots, \bs{p}'_{n-1})
\Big].
\end{align}
Using \eqref{trick1} for all the $\K^\kappa(\Delta_m^{(n)};\bs{p}_m)$ terms with $m=1,\ldots,n-1$, this becomes
\begin{align}
\mathcal{E}^{(n)\kappa}_{SCWI}&
=\prod_{i = 1}^{n-1} C_{in} \int \frac{\D^d \bs{q}_{i}}{(2 \pi)^d} \frac{1}{q_1^{2 \alpha_{1n} + d} q_2^{2 \alpha_{2n} + d} \ldots q_{n-1}^{2 \alpha_{n-1,n} + d}}\nn\\&\quad\times\Big[M_{n-1}(\bs{p}'_1, \ldots, \bs{p}'_{n-1}) \Big(\K^\kappa(\Delta_n^{(n)};\bs{p}_n)+\sum_{m=1}^{n-1}\K^\kappa(\alpha_{mn}+d;\bs{q}_m)\Big)(2 \pi)^d \delta( \bs{p}'_n) \nn\\&\quad
+(2 \pi)^d \delta( \bs{p}'_n)
\Big(\sum_{m=1}^{n-1} \K^\kappa(\Delta_m^{(n)}+\alpha_{mn};\bs{p}'_m) \Big)M_{n-1}(\bs{p}'_1, \ldots, \bs{p}'_{n-1})
\Big].
\end{align}
However, the first set of terms acting on the delta function vanishes, since from the definition \eqref{Kopdef} we have
\begin{align}
&\Big(\K^\kappa(\Delta_n^{(n)};\bs{p}_n)+\sum_{m=1}^{n-1}\K^\kappa(\alpha_{mn}+d;\bs{q}_m)\Big)(2 \pi)^d \delta( \bs{p}'_n) \nn\\
&\quad =
\K^\kappa(\Delta_n^{(n)}+\sum_{m=1}^{n-1}\alpha_{mn};\,\bs{p}'_n)(2 \pi)^d \delta( \bs{p}'_n) \nn\\&
\quad =
\K^\kappa(0;\bs{p}'_n)(2 \pi)^d \delta( \bs{p}'_n) \nn\\&
\quad =0.
\end{align}
To obtain the third line here we used \eqref{deltaijcond_a}, while the final line follows from \eqref{trick2}.  This leaves us with
\begin{align}
\mathcal{E}^{(n)\kappa}_{SCWI}&
=\prod_{i = 1}^{n-1} C_{in} \int \frac{\D^d \bs{q}_{i}}{(2 \pi)^d} \frac{1}{q_1^{2 \alpha_{1n} + d} q_2^{2 \alpha_{2n} + d} \ldots q_{n-1}^{2 \alpha_{n-1,n} + d}}\nn\\&\quad\times(2 \pi)^d \delta( \bs{p}'_n)
\Big(\sum_{m=1}^{n-1} \K^\kappa(\Delta_m^{(n-1)};\bs{p}'_m) \Big)M_{n-1}(\bs{p}'_1, \ldots, \bs{p}'_{n-1})
\end{align}
where we used  \eqref{deltainc} to replace  $\Delta_m^{(n)}+\alpha_{mn} = \Delta_m^{(n-1)}$ in the operator arguments.  The last line is however just the special conformal Ward identity for the $(n-1)$-point function:
\begin{align}
\mathcal{E}^{(n)\kappa}_{SCWI}&
=\prod_{i = 1}^{n-1} C_{in} \int \frac{\D^d \bs{q}_{i}}{(2 \pi)^d} \frac{1}{q_1^{2 \alpha_{1n} + d} q_2^{2 \alpha_{2n} + d} \ldots q_{n-1}^{2 \alpha_{n-1,n} + d}}(2 \pi)^d \delta( \bs{p}'_n)\,
\mathcal{E}^{(n-1)\kappa}_{SCWI}.
\end{align}
Thus, if the special conformal Ward identity for the $(n-1)$-point function is satisfied, that for the $n$-point function is too. Since \eqref{trick2} proves the Ward identity for $n=1$, it then follows for all $n \geq 1$ by recursion.

\subsection{Total derivatives for mesh integrals} \label{app:conf_mesh_A}

Here we want to prove that \eqref{to_show_mesh} holds, establishing that the action of the special conformal Ward identity on the mesh integral is a total derivative. To simplify the expressions involved, we multiply both sides by $\text{Den}_n(\bs{\alpha})$ after which we wish to prove that
\begin{align} \label{to_show_mesh_a}
& \sum_{m=1}^{n-1} \text{Den}_n(\bs{\alpha}) \mathcal{K}^{\kappa}(\Delta_m; \bs{p}_m) \left[ \frac{1}{\text{Den}_n(\bs{\alpha})} \right] =  \sum_{\substack{i,j = 1\\i \neq j}}^{n-1} \text{Den}_n(\bs{\alpha}) \frac{\partial}{\partial q_{ij}^\mu} \left[ (2 \alpha_{in} + d) \frac{A_{ij}^{\kappa \mu}}{\text{Den}_n(\bs{\alpha})} \right].
\end{align}
This is essentially a long but straightforward calculation. Using
\begin{align}
\frac{\partial}{\partial p^\mu} \frac{1}{(\bs{l} - \bs{p})^{2 \gamma}} & = 2 \gamma \frac{(\bs{l} - \bs{p})_\mu}{(\bs{l} - \bs{p})^{2 \gamma + 2}}
\end{align}
we find that each element under the sum on the left-hand side of \eqref{to_show_mesh_a} is
\begin{align}
& \text{Den}_n(\bs{\alpha}) \mathcal{K}^{\kappa}(\Delta_m; \bs{p}_m) \left[ \frac{1}{\text{Den}_n(\bs{\alpha})} \right]  \nn\\
& \qquad = - 2 (2 \alpha_{mn} + d)(2 \alpha_{mn} + d + 2) \frac{(\bs{l}_m - \bs{p}_m)^\kappa}{(\bs{l}_m - \bs{p}_m)^{4}} \bs{l}_m \cdot (\bs{l}_m - \bs{p}_m) \nn\\
& \qquad\quad + 2 (2 \alpha_{mn} + d) (\alpha_{mn} + 2) \frac{l_m^\kappa}{(\bs{l}_m - \bs{p}_m)^{2}} \nn\\
& \qquad\quad + 2 (2 \alpha_{mn} + d) (\alpha_{mn} + \Delta_m) \frac{(\bs{l}_m - \bs{p}_m)^\kappa}{(\bs{l}_m - \bs{p}_m)^{2}}. \label{res1}
\end{align}
On the right-hand side of \eqref{to_show_mesh_a}, notice that the vector $\bs{q}_{ij}$ appears three times within  $\text{Den}_n(\bs{\alpha})$ in \eqref{Den}: once as an explicit factor of $\bs{q}_{ij}$, and then implicitly as a term $+\bs{q}_{ij}$ within 
$\bs{l}_i$, and a term $-\bs{q}_{ij}$ in $\bs{l}_j$. Using
\begin{align}
\frac{\partial}{\partial q_{ij}^\mu} \bs{l}_m^\nu & = \delta^\nu_\mu ( \delta_{im} - \delta_{jm} ),
\end{align}
we find each element under the double sum on the right-hand side of \eqref{to_show_mesh_a} is
\begin{align}
 \text{Den}_n(\bs{\alpha}) \frac{\partial}{\partial q_{ij}^\mu} \left[ \frac{A_{ij}^{\kappa \mu}}{\text{Den}_n(\bs{\alpha})} \right]  
&= - 2 (2 \alpha_{in} + d + 2) \frac{(\bs{l}_i - \bs{p}_i)^\kappa}{(\bs{l}_i - \bs{p}_i)^{4}} \bs{q}_{ij} \cdot (\bs{l}_i - \bs{p}_i) \nn\\
& \quad + 2 (\alpha_{in} + 2) \frac{q_{ij}^\kappa}{(\bs{l}_i - \bs{p}_i)^2} - 2 \alpha_{ij} \frac{(\bs{l}_i - \bs{p}_i)^\kappa}{(\bs{l}_i - \bs{p}_i)^2}.
\end{align}
Now the sum over $j$ can be carried out and the remaining index $i$ re-labelled as $m$.  For use in the next section, we label this result as 
\begin{align}
\mathcal{R}_m^{\kappa}(\bs{\alpha}) & = \text{Den}_n(\bs{\alpha}) \sum_{\substack{j = 1\\j \neq m}}^{n-1} \frac{\partial}{\partial q_{mj}^\mu} \left[  \frac{A_{mj}^{\kappa \mu}}{\text{Den}_n(\bs{\alpha})} \right] \nn\\
& = - 2 (2 \alpha_{mn} + d + 2) \frac{(\bs{l}_m - \bs{p}_m)^\kappa}{(\bs{l}_m - \bs{p}_m)^{4}} \bs{l}_m \cdot (\bs{l}_m - \bs{p}_m) \nn\\[0.5ex]
& \quad + 2  (\alpha_{mn} + 2) \frac{l_m^\kappa}{(\bs{l}_m - \bs{p}_m)^{2}} 
- 2  ( \alpha^t_{m} - \alpha_{mn} ) \frac{(\bs{l}_m - \bs{p}_m)^\kappa}{(\bs{l}_m - \bs{p}_m)^{2}}, \label{Rdef}
\end{align}
where $\alpha^t_m = \sum_{j=1}^n \alpha_{mj}$. When multiplied by $(2 \alpha_{mn} + d)$, this expression matches \eqref{res1} provided the  conditions \eqref{deltaijcond_a} hold.

\subsection{Explicit total derivatives for simplex integrals} \label{app:conf_simp_A}

In this appendix, we prove by direct calculation that the action of the special conformal Ward identity on the simplex is a total derivative of the form \eqref{to_show_simp_k}, with  coefficients given by \eqref{G0} and \eqref{GI}.
To begin with, 
as in appendix \ref{app:conf_mesh_A}, we simplify the algebra by multiplying both sides of \eqref{to_show_simp_k} by $\text{Den}_n(\bs{\alpha})$. Furthermore, we reshuffle the indices on any cross ratio $\hat{u}_{[pqrs]}$ so that $s$ is the largest index. This can always be done, since by definition the cross ratios \eqref{conf_ratio_q} have the $\Z_2 \times \Z_2$ redundancy
\begin{align} \label{red_conf_ratio}
\hat{u}_{[pqrs]} = \hat{u}_{[qpsr]} = \hat{u}_{[rspq]} = \hat{u}_{[srqp]}.
\end{align}
With this convention, only the final  index $s$ can be equal to $n$ meaning \eqref{to_show_simp_k}  reads
\begin{align} \label{to_show_simp_a}
& \sum_{m=1}^{n-1} \text{Den}_n(\bs{\alpha}) \mathcal{K}^{\kappa}(\Delta_m; \bs{p}_m) \left[ \frac{\hat{f}(\hat{\bs{u}})}{\text{Den}_n(\bs{\alpha})} \right] = \sum_{\substack{i,j = 1\\i \neq j}}^{n-1} \text{Den}_n(\bs{\alpha}) \frac{\partial}{\partial q_{ij}^\mu} \left[ (2 \alpha_{in} + d) \frac{A_{ij}^{\kappa \mu}}{\text{Den}_n(\bs{\alpha})} \hat{f}(\hat{\bs{u}}) \right. \nn\\
& \qquad\qquad\qquad\qquad \left. + \sum_{[pqrn] \in \mathcal{U}} 2 (\delta_{iq} - \delta_{ir}) \frac{A^{\kappa \mu}_{ij} \hat{u}_{[pqrn]}}{\text{Den}_n(\bs{\alpha})} \frac{\partial \hat{f}(\hat{\bs{u}})}{\partial \hat{u}_{[pqrn]}} \right].
\end{align}
Here we substituted in \eqref{G0} and \eqref{GI}, and the final 
sum is taken over the set of indices of the form $[pqrn]$.
In the following, we will now demonstrate this equation is indeed correct.

We begin with the left-hand side of \eqref{to_show_simp_a}, and use the fact that
\begin{align}
\frac{\partial \hat{u}_{[pqrs]}}{\partial p_m^\mu} & = \frac{2 (\bs{l}_m - \bs{p}_m)_{\mu}}{(\bs{l}_m - \bs{p}_m)^2} \hat{u}_{[pqrs]} \, \delta_{ns} (\delta_{mq} - \delta_{mr}), \nn\\[1ex]
\frac{\partial^2 \hat{u}_{[pqrs]}}{\partial p_m^\mu \partial p_m^\nu} & = \left[ \frac{8 (\bs{l}_m - \bs{p}_m)_{\mu} (\bs{l}_m - \bs{p}_m)_{\nu}}{(\bs{l}_m - \bs{p}_m)^4} - \frac{2 \delta_{\mu\nu}}{(\bs{l}_m - \bs{p}_m)^2} \right] \hat{u}_{[pqrs]} \delta_{ns} (\delta_{mq} - \delta_{mr}).
\end{align}
These expressions are non-vanishing only if $s = n$. The left-hand side of \eqref{to_show_simp_a} now takes the form
\begin{align} \label{res3}
& \text{Den}_n(\bs{\alpha}) \mathcal{K}^{\kappa}(\Delta_m; \bs{p}_m) \left[ \frac{\hat{f}(\hat{\bs{u}})}{\text{Den}_n(\bs{\alpha})} \right] 
\nn\\[0.5ex]&
\quad = \mathcal{A}_{m}^{\kappa} \hat{f}(\hat{\bs{u}}) \nn\\[0.5ex]
& \qquad + \sum_{[pqrn] \in \mathcal{U}} \frac{\partial \hat{f}(\hat{\bs{u}})}{\partial \hat{u}_{[pqrn]}} \hat{u}_{[pqrn]} (\delta_{mq}-\delta_{mr}) \mathcal{B}_{m}^{\kappa} \nn\\
& \qquad + \sum_{\substack{[pqrn] \in \mathcal{U}\\ [p' q' r' n] \in \mathcal{U}}} 
\frac{\partial^2 \hat{f}(\hat{\bs{u}})}{\partial \hat{u}_{[pqrn]} \partial \hat{u}_{[p'q'r'n]}} \hat{u}_{[pqrn]} \hat{u}_{[p'q'r'n]} (\delta_{mq}-\delta_{mr}) (\delta_{mq'}-\delta_{mr'}) \mathcal{C}_{m}^{\kappa},
\end{align}
where $\mathcal{A}_{m}^{\kappa}$ is given by $\eqref{res1}$ and
\begin{align}
\mathcal{B}_{m}^{\kappa} & = - 8 (2 \alpha_{mn} + d + 2) \frac{(\bs{l}_m - \bs{p}_m)^\kappa}{(\bs{l}_m - \bs{p}_m)^{4}} \bs{l}_m \cdot (\bs{l}_m - \bs{p}_m) \nn\\
& \qquad\qquad + 2 (4 \alpha_{mn} + d + 6) \frac{l_m^\kappa}{(\bs{l}_m - \bs{p}_m)^{2}} \nn\\
& \qquad\qquad + \: 2 (4 \alpha_{mn} + 2 \Delta_m + d + 2) \frac{(\bs{l}_m - \bs{p}_m)^\kappa}{(\bs{l}_m - \bs{p}_m)^{2}}, \\[1ex]
\mathcal{C}_{m}^{\kappa} & = - 8 \frac{(\bs{l}_m - \bs{p}_m)^\kappa}{(\bs{l}_m - \bs{p}_m)^{4}} \bs{l}_m \cdot (\bs{l}_m - \bs{p}_m) \nn\\
& \qquad\qquad + \: 4 \frac{l_m^\kappa}{(\bs{l}_m - \bs{p}_m)^{2}} + 4 \frac{(\bs{l}_m - \bs{p}_m)^\kappa}{(\bs{l}_m - \bs{p}_m)^{2}}.
\end{align}

Next we analyse the right-hand side of \eqref{to_show_simp_a}. We find:
\begin{align} \label{res4}
& \left[ \text{rhs of } \eqref{to_show_simp_a} \right] = \sum_{m=1}^{n-1} (2 \alpha_{mn} + d) \mathcal{R}^\kappa_m(\bs{\alpha}) \hat{f}(\hat{\bs{u}}) \nn\\
& \qquad\qquad + \sum_{m=1}^{n-1} \sum_{[pqrn] \in \mathcal{U}} \Big[ (2 \alpha_{mn,[pqrn]} + d) \mathcal{R}^{\kappa}_m(\bs{\alpha}_{[pqrn]}) \nn\\
& \qquad\qquad\qquad\qquad  - \: (2 \alpha_{mn} + d) \mathcal{R}^\kappa_m(\bs{\alpha}) \Big] \hat{u}_{[pqrn]} \frac{\partial \hat{f}(\hat{\bs{u}})}{\partial \hat{u}_{[pqrn]}} \nn\\
& \qquad \qquad+ \sum_{\substack{i,j = 1\\i \neq j}}^{n-1}\sum_{\substack{[pqrn] \in \mathcal{U}\\ [p' q' r' s'] \in \mathcal{U}}} 2 (\delta_{iq} - \delta_{ir}) A^{\kappa \mu}_{ij} \hat{u}_{[pqrn]} \frac{\partial \hat{u}_{[p'q'r's']}}{\partial q_{ij}^{\mu}} \frac{\partial^2 \hat{f}(\hat{\bs{u}})}{\partial \hat{u}_{[pqrn]} \partial \hat{u}_{[p'q'r's']}}.
\end{align}
The expression for $\mathcal{R}^\kappa_m(\bs{\alpha})$ was given in \eqref{Rdef}, and we also used \eqref{simplex_to_mesh}. The parameters $\bs{\alpha}_{[pqrs]}$ are given by \eqref{a_abcd} with $\gamma_1 = 1$, namely
\begin{align} \label{a_abcd_a}
\alpha_{ij, [pqrs]} = \alpha^{(1)}_{ij, [pqrs]} = \alpha_{ij} + \delta_{ip} \delta_{jr} + \delta_{iq} \delta_{js} - \delta_{ip} \delta_{jq} - \delta_{ir} \delta_{js}.
\end{align}
In the last line of \eqref{res4} we have an index $s'$ not equal to $n$, but we still use \eqref{red_conf_ratio} so that $p',q',r' < s'$. The first term reproduces $\mathcal{A}_m^{\kappa} \hat{f}$. The difference in the second term reads
\begin{align}
& (2 \alpha_{mn,[pqrn]} + d) \mathcal{R}^{\kappa}_m(\bs{\alpha}_{[pqrn]}) - (2 \alpha_{mn} + d) \mathcal{R}^\kappa_m(\bs{\alpha}) = \nn\\
& \qquad = - 8 (2 \alpha_{mn} + d + 2) (\delta_{mq}-\delta_{mr}) \frac{(\bs{l}_m - \bs{p}_m)^\kappa}{(\bs{l}_m - \bs{p}_m)^{4}} \bs{l}_m \cdot (\bs{l}_m - \bs{p}_m) \nn\\
& \qquad\qquad + 2 (4 \alpha_{mn} + d + 6) (\delta_{mq}-\delta_{mr}) \frac{l_m^\kappa}{(\bs{l}_m - \bs{p}_m)^{2}} \nn\\
& \qquad\qquad + 2 (4 \alpha_{mn} - 2 \alpha^t_{m} + d + 2) (\delta_{mq}-\delta_{mr}) \frac{(\bs{l}_m - \bs{p}_m)^\kappa}{(\bs{l}_m - \bs{p}_m)^{2}},
\end{align}
where for the last line we use the fact that $\alpha^t_m$ is the same for $\mathcal{R}^{\kappa}_m(\bs{\alpha}_{pqrn})$ and $\mathcal{R}^{\kappa}_m(\bs{\alpha})$. Clearly, this reproduces the second term in \eqref{res3} provided the condition \eqref{deltaijcond_a} holds.

For the last term in \eqref{res4}, we use the fact that for $i < j$ we have
\begin{align}
\frac{\partial \hat{u}_{[pqrs]}}{\partial q_{ij}^\mu} & = \frac{2 q_{ij \mu}}{q_{ij}^2} \hat{u}_{[pqrs]} \left[ \delta_{ip} \delta_{jq} + \delta_{ir} \delta_{js} - \delta_{ip} \delta_{jr} - \delta_{iq} \delta_{js} \right], \quad s < n, \\
\frac{\partial \hat{u}_{[pqrn]}}{\partial q_{ij}^\mu} & = \frac{2 q_{ij \mu}}{q_{ij}^2} \hat{u}_{[pqrn]} \left[ \delta_{ip} \delta_{jq} - \delta_{ip} \delta_{jr} \right] - \frac{2(\bs{l}_i - \bs{p}_i)_{\mu}}{(\bs{l}_i - \bs{p}_i)^2} \hat{u}_{[pqrn]} (\delta_{iq} - \delta_{ir}) \nn\\
& \qquad\qquad + \frac{2(\bs{l}_j - \bs{p}_j)_{\mu}}{(\bs{l}_j - \bs{p}_j)^2} \hat{u}_{[pqrn]} (\delta_{jq} - \delta_{jr}).
\end{align}
Next we split the first sum in the last term of \eqref{res4} into sums over $i < j$ and $i > j$, and the inner sum into two cases: $s' < n$ and $s' = n$. It is easy to see that as far as the sum over $s' < n$ is concerned, the sums over $i < j$ and $i > j$ cancel. After the dust settles, the remaining sum matches the last line of \eqref{res3} exactly,  concluding the proof of \eqref{to_show_simp_a}.

\section{The Symanzik trick for conformal integrals}
\label{Symanzik_trick_appendix}

The position-space contact Witten diagram for a  holographic CFT is proportional to the conformal star integral of Symanzik  
\cite{Symanzik:1972wj}.\footnote{Note however this conformal star lives 
in a spacetime dimension $\Delta_t$, which is not necessarily equal to the dimension $d$ of the holographic CFT.}  In particular,  Symanzik showed  that the Schwinger parametrisation for this integral
 \begin{align}\label{startingpoint}
S_n &= \Big(\prod_{i=1}^n \int_0^\infty \D s_i\, s_i^{\Delta_i-1} \Big) s_t^{-\Delta_t/2}  e^{-\frac{1}{s_t}\sum_{i<j}s_is_j x_{ij}^2}
\end{align}
is unchanged by making the replacement
\[\label{Symsub}
s_t =\sum_{i=1}^n s_i \rightarrow \sum_{i=1}^n \kappa_i s_i
\]
for any set of constants $\kappa_i\ge 0$ not all zero. 
While elementary, a derivation of this result is perhaps still useful and so 
 we present one here, elaborating on  appendix B of \cite{Dolan:2000uw}.

The calculation proceeds by changing variables from the $s_i$ to the set $(\sigma,y_i)$, where
\[\label{svars}
s_i=\sigma y_i, \qquad \sum_{i=1}^n \kappa_i y_i=1.
\]
The Jacobian for this transformation can be obtained either by inspection (as below), or else through use of the Schur complement identity
\begin{align}
J = \left(\begin{array}{cc}
A & B\\ C & D
\end{array}\right) & = 
\left(\begin{array}{cc}
I\,\, & B D^{-1}\\ 0 & I
\end{array}\right) 
\left(\begin{array}{cc}
A -B D^{-1}C\, & 0\\ 0 & D
\end{array}\right) 
\left(\begin{array}{cc}
I & 0\\ D^{-1}C\, & I
\end{array}\right), 
\end{align}
which implies
\[
\det{J} = \det{D}\det{(A-B D^{-1}C)}.
\]
Explicitly, setting $y_n = \kappa_n^{-1}(1-\sum_{i=1}^{n-1}\kappa_i y_i)$ and transposing a column for convenience, 
\begin{align}
\left|
\begin{array}{cccc}
\dfrac{\partial s_n}{\partial \sigma} & 
\dfrac{\partial s_1}{\partial \sigma} & 
\ldots & 
\dfrac{\partial s_{n-1}}{\partial \sigma}  
\\ [2ex]
\dfrac{\partial s_n}{\partial y_1}&
\dfrac{\partial s_1}{\partial y_1}& 
\ldots &
\dfrac{\partial s_{n-1}}{\partial y_1}  \\[2ex]
\vdots & \vdots &\ddots &\vdots \\[2ex]
\dfrac{\partial s_n}{\partial y_{n-1}}&
\dfrac{\partial s_1}{\partial y_{n-1}}& 
\ldots &
\dfrac{\partial s_{n-1}}{\partial y_{n-1}} 
\end{array}
\right|
= 
\left|
\begin{array}{cccc}
y_n& 
y_1& 
\ldots & 
y_{n-1}
\\ [2ex]
-\sigma \kappa_1/\kappa_n&
\sigma & 
\ldots &
0 \\[2ex]
\vdots & \vdots &\ddots &\vdots \\[2ex]
-\sigma\kappa_{n-1}/\kappa_n&
0& 
\ldots &
\sigma
\end{array}
\right|. \label{Jdet}
\end{align}
Taking
\[
A = y_n, \quad B = (y_1, \ldots, y_{n-1}), \quad C^T =-\sigma \kappa_n^{-1} (\kappa_1, \ldots,\kappa_{n-1}), \quad D = \sigma I,
\]
we then have
\[
\det{J} = \sigma^{n-1} (y_n +\kappa_n^{-1} \sum_{i=1}^{n-1}y_i  \kappa_i) = \kappa_n^{-1}\sigma^{n-1}.
\]
Alternatively, the determinant in \eqref{Jdet} can be evaluated by inspection after observing that,  after elimination of the first row and the $i$th column, for $i>1$ the only the first entry on the $(i-1)$-th row of the resulting subdeterminant is nonzero.

The change of measure is therefore 
\[
\prod_{i=1}^n \D s_i = \kappa_n^{-1}  \sigma^{n-1} \D\sigma \prod_{i=1}^{n-1}\D y_i =  \sigma^{n-1} \D\sigma \prod_{i=1}^{n}\D y_i \,\delta(1-\sum_{i=1}^n\kappa_i y_i).
\]
In these new variables, the integral \eqref{startingpoint} is now
 \begin{align}
S_n &= \Big(\prod_{i=1}^n \int_0^1 \D y_i\, y_i^{\Delta_i-1} \Big)\delta\big(1-\sum_{i=1}^n \kappa_i y_i\big) y_t^{-\Delta_t/2} \int_0^\infty\D\sigma\,\sigma^{\Delta_t/2-1} e^{-\frac{\sigma}{y_t}\sum_{i<j}y_i y_j x_{ij}^2}.
\end{align}
Evaluating the $\sigma$ integral, the crucial point is that  the factor of $y_t^{-\Delta_t/2}$ now cancels out:
 \begin{align}\label{Sym_res1}
S_n &= \Gamma\big(\frac{\Delta_t}{2}\big)\Big(\prod_{i=1}^n \int_0^1 \D y_i\, y_i^{\Delta_i-1} \Big)\delta\big(1-\sum_{i=1}^n \kappa_i y_i\big) \Big(\sum_{i<j} y_i y_j x_{ij}^2\Big)^{-\Delta_t/2}.
\end{align}
As a result, one would also have arrived at this same expression had we started instead from \eqref{startingpoint} with the substitution \eqref{Symsub}.   This substitution therefore leaves the value of the integral unchanged.

\section{\texorpdfstring{The holographic $D$-function as a  triple-$K$ integral}{The holographic D-function as  triple-K integral}}\label{app_Dfn}

The  4-point contact Witten diagram in position space, often referred to as the holographic `$D$-function' \cite{DHoker:1999kzh}, is equivalent to a conformal  Symanzik star  integral living in spacetime dimension $\Delta_t$ \cite{Symanzik:1972wj, Dolan:2000ut}.  Converting to the dual conformal (or region) momenta, this in turn becomes a conformal box integral.  In this appendix, we point out that the $D$-function can also be re-written as a 1-loop triangle or triple-$K$ integral.   This follows either from the analysis of section \ref{recursive_sec}, or by direct inversion as we discuss here.
A connection between these integrals can be anticipated given they all have an explicit evaluation in terms of Appell $F_4$ \cite{Boos:1990rg, Davydychev:1992xr, Dolan:2000uw, Bzowski:2013sza}.

The $D$-function, corresponding to  $\mathcal{I}_4$ in \eqref{posspcontact}, is equal to the conformal star integral
\[\
\mathcal{I}_4 = c\int\D^{\Delta_t}\x_5 \frac{1}{x_{15}^{2\Delta_1}x_{25}^{2\Delta_2}x_{35}^{2\Delta_3}x_{45}^{2\Delta_4}},
\]
where 
\[
c = \frac{1}{2}\pi^{-(\Delta_t+3d)/2}\Gamma\Big(\frac{\Delta_t-d}{2}\Big)\prod_{i=1}^4\frac{\Gamma(\Delta_i)}{\Gamma(\Delta_i-d/2)}.
\]
To verify this, one Schwinger parametrises the propagators then performs the integral over $\x_5$, making use of \eqref{compsq1}, to arrive at \eqref{step1}.
This conformal star integral can be converted to a conformal box by introducing the dual momenta
\begin{align}
&\bs{k}_1=\bs{x}_{41}, \quad \bs{k}_2=\bs{x}_{12}, \quad \bs{k}_3=\bs{x}_{23},\quad \bs{k}_4=\bs{x}_{34}, \quad \bs{k}=\bs{x}_{54},
\end{align}
and defining
\[
\bs{K}_i = \sum_{j=1}^i \bs{k}_j \qquad \Rightarrow \qquad
\bs{K}_1 = \bs{x}_{41}, \qquad \bs{K}_2 = \bs{x}_{42}, \qquad \bs{K}_3=\bs{x}_{43},
\]
whereupon
\begin{align}
\mathcal{I}_4 = c \int \D^{\Delta_t}\bs{k}\,\frac{1}{\vphantom{\sum^b}|\bs{k}+\bs{K}_1|^{2\Delta_1}|\bs{k}+\bs{K}_2|^{2\Delta_2}|\bs{k}+\bs{K}_3|^{2\Delta_3}|\bs{k}|^{2\Delta_4}}.
\end{align}
We can now eliminate a propagator by inverting.  Defining
\[
\bs{k} = \frac{\bs{q}}{q^2}, \qquad \bs{K}_i = \frac{\bs{Q}_i}{Q_i^2},
\]
we obtain the 1-loop triangle integral
\begin{align}
\mathcal{I}_4 = c \,Q_1^{2\Delta_1}Q_2^{2\Delta_2}Q_3^{2\Delta_3}
 \int \D^{\Delta_t}\bs{q}\,\frac{1}{\vphantom{\sum^b}|\bs{q}+\bs{Q}_1|^{2\Delta_1}|\bs{q}+\bs{Q}_2|^{2\Delta_2}|\bs{q}+\bs{Q}_3|^{2\Delta_3}},
\end{align}
where the total dimension $\Delta_t =\sum_{i=1}^4\Delta_i$ is still that of the 4-point function.  
This triangle integral can now be converted to a triple-$K$ integral using the results of section \ref{startrisection}, or equivalently appendix A.3 in \cite{Bzowski:2013sza}.  This yields 
\begin{align}
\mathcal{I}_4 &= \frac{c \,2^4 (\pi/2)^{\Delta_t/2}}{\prod_{i=1}^4\Gamma(\Delta_i)}  \,Q_1^{2\Delta_1}Q_2^{2\Delta_2}Q_3^{2\Delta_3}\nn\\& \quad\times
I_{\Delta_t/2-1, \, \{\Delta_1+\Delta_4-\Delta_t/2,\Delta_2+\Delta_4-\Delta_t/2,\Delta_3+\Delta_4-\Delta_t/2\}}(|\bs{Q}_3-\bs{Q}_2|,|\bs{Q}_1-\bs{Q}_3|,|\bs{Q}_2-\bs{Q}_1|),
\end{align}
where the triple-$K$ integral is defined in \eqref{tripleKdef}.
 The squared magnitudes of the momenta appearing here are
\begin{align}
|\bs{Q}_3-\bs{Q}_2|^2 &= \left|\frac{\bs{K}_3}{K_3^2}-\frac{\bs{K}_2}{K_2^2}\right|^2 = \frac{1}{x_{34}^2}+\frac{1}{x_{24}^2}-\frac{2\bs{x}_{43}\cdot\bs{x}_{42}}{x_{34}^2x_{24}^2}=\frac{(\bs{x}_{43}-\bs{x}_{42})^2}{x_{34}^2x_{24}^2}=\frac{x_{23}^2}{x_{34}^2x_{24}^2}, \\
|\bs{Q}_1-\bs{Q}_3|^2 &= \frac{x_{13}^2}{x_{14}^2x_{34}^2},\\
|\bs{Q}_2-\bs{Q}_1|^2 &=  \frac{x_{12}^2}{x_{14}^2x_{24}^2}.
\end{align}
To rewrite this result in terms of cross ratios, we rescale the integration variable $z$ in the triple-$K$ integral as $z\rightarrow z\sqrt{x_{14}^2x_{34}^2/x_{13}^2}$.
We now obtain
\[\label{fuvres0}
\mathcal{I}_4 = \prod_{i<j} x_{ij}^{2\alpha_{ij}} f(u,v)
\]
where 
\begin{align}\label{fuvres1}
f(u,v)&= \frac{2^4 (\pi/2)^{\Delta_t/2}}{\prod_{i=1}^4\Gamma(\Delta_i)} \,c\,u^{-\alpha_{14}} v^{\alpha_{34}}  I_{\Delta_t/2-1, \, \{\alpha_{14}-\alpha_{23},\,\alpha_{24}-\alpha_{13},\,\alpha_{34}-\alpha_{12}\}}(\sqrt{u},1,1/\sqrt{v})
\end{align}
and
 \[\label{uvdef}
 u = \frac{x_{14}^2 x_{23}^2}{x_{13}^2x_{24}^2}, \qquad v = \frac{x_{13}^2 x_{24}^2}{x_{12}^2x_{34}^2}, \qquad 2\alpha_{ij}=\Delta_t/3-\Delta_i-\Delta_j.
 \]
 Equations \eqref{fuvres0} and \eqref{fuvres1} thus
express the $D$-function or position-space contact 4-point function as a triple-$K$ integral of the cross ratios.
 We have chosen this specific parametrisation of the cross ratios so that upon substituting $x_{ij}^2\rightarrow q_{ij}^2$ they coincide with our choice of momentum-space cross ratios 
 $\hat{u}$ and $\hat{v}$ in \eqref{uvhatdef}.
The resemblance of \eqref{fuvres1} to \eqref{fhatres} reflects the fact that $f_n(\bs{u})$ and $\hat{f}_n(\hat{\bs{u}})$ for the contact diagram have the same functional form up to  a change of parameters, as discussed at the end of section \ref{recursive_sec}.

\bibliographystyle{JHEP}
\bibliography{long}

\end{document}